# Brownian spin-locking effect

Xiao Zhang[1]†, Peiyang Chen[1,2]†, Mei Li[1], Yuzhi Shi[3], Erez Hasman[4]*, Bo Wang[1]*, Xianfeng Chen[1,5,6]*

[1]State Key Laboratory of Advanced Optical Communication Systems and Networks, School of Physics and Astronomy, Shanghai Jiao Tong University; Shanghai, 200240, China.
[2]Zhiyuan College, Shanghai Jiao Tong University, Shanghai, 200240, China.
[3]Institute of Precision Optical Engineering, School of Physics Science and Engineering, Tongji University, Shanghai 200092, China.
[4]Atomic-Scale Photonics Laboratory, Russell Berrie Nanotechnology Institute, and Helen Diller Quantum Center, Technion – Israel Institute of Technology; Haifa, 3200003, Israel.
[5]Shanghai Research Center for Quantum Sciences; Shanghai, 201315, China.
[6]Collaborative Innovation Center of Light Manipulations and Applications, Shandong Normal University; Jinan, 250358, China.

*Corresponding authors:
mehasman@technion.ac.il (E.H.)
wangbo89@sjtu.edu.cn (B.W.)
xfchen@sjtu.edu.cn (X.C.)

**Abstract**
Brownian systems are characterized by spatiotemporal disorder, which arises from the erratic motion of particles driven by thermal fluctuations. When light interacts with such systems, it typically produces unpolarized and uncorrelated fields. Here, we report the observation of a large-scale spin-locking effect of light within a Brownian medium. In an observation direction perpendicular to the incident wave's momentum, scattering naturally divides into two diffusion regions, each associated with an opposite spin from the Brownian nanoparticles. This effect arises from the intrinsic spin-orbit interactions of scattering from individual nanoparticles, which ubiquitously generate radiative spin fields that propagate through the Brownian medium with multiple incoherent scattering. It offers a novel experimental platform for exploring macroscale spin behaviors of diffused light, with potential applications in precision metrology for measuring various nanoparticle properties. Our findings may inspire the study of analogous phenomena for different waves from novel spin-orbit interactions in complex disordered systems.

**Introduction**
The study of coherent wave-interaction with complex disordered structures has led to many unique wave phenomena, such as Anderson localization (*1, 2*), super oscillation (*3*), branched flow (*4*), and Hall effects (*5*). Spin is an intrinsic angular momentum carried by different waves. It plays a central role in both classical and quantum phenomena and has broad implications in various branches of physics. In condensed matter physics, spin interactions give rise to exotic states of matter, such as topological insulators (*6, 7*) and spin liquids (*8, 9*), while magnetic properties emerge due to the coupling of spin and orbital degrees of freedom (*10*). In optics, spin angular momentum is associated with light's circular polarization. The interplay between light and matters gives rise to many optical spin-dependent effects, such as Rashba effect (*11*), spin Hall effect (*12*), and universal spin-momentum locking of evanescent waves (*13-15*). Many spin-dependent phenomena are described as stable separation between two spin components of light, which can be characterized by a geometric-phase gradient $\partial\Phi_G/\partial\xi$ that occurs in a general parameter space $\xi$, driven by various spin-orbit interactions (SOIs) (*16*). Generally, $\partial\Phi_G/\partial\xi$ arises from spatial asymmetric light-matter interactions that are either induced from asymmetric optical illuminations (*17, 18*) or engineered structures (*19-21*). However, coherent waves can severely destroy SOI from an overwhelming randomness. As the scale of disorder increases, a delicate symmetry-breaking condition becomes less defined, eliminating spin split phenomena as $<\partial\Phi_G/\partial\xi>$ approaches zero (*22, 23*).

Compared to static disordered structures, a Brownian system is not only spatially disordered but also temporally fluctuated. Brownian motions exist widely in nature. It is not only a cornerstone of statistical physics (*24*), but also has numerous practical applications in a wide range of research fields including chemistry, biology, and even finance. In optics, the diffusion properties of Brownian particles result in an optical technique to measure the size of nanoparticles, known as dynamical light scattering (DLS) (*25*). For colloidal suspensions, the random movement of particles are resulted from their stochastic collisions with fluid molecules. The locations of these Brownian particles are therefore unpredictable, and the mean squared displacements of these particles are proportional to time.

Here, we report an unforeseen Brownian spin-locking effect (BSLE). This phenomenon arises from the spatiotemporally disordered scattering from a massive number of spherical nanoparticles in the colloidal suspension. Instead of being destroyed from strong disorder, the spin fields are divided into two opposite macroscale regions. Its physical mechanism is composed of two aspects, one is the intrinsic SOI from nanoparticles, and the other is the incoherent multiple scattering from these nanoparticles. Firstly, spin symmetry broken emerges in an observation direction perpendicular to the incident momentum. This is a universal effect for optical scattering that does not require any structural anisotropy or chirality. Specifically, it can emerge for isotropic spherical nanoparticles excited by linearly polarized plane waves. The scattered spin is in general partially aligned with the Poynting vector, and it can be characterized by a power-law decaying function ($1/r^2$) and radiates to far field to obtain multiple scattering. Secondly, the erratic movement of many nanoparticles leads to constant density variation and incoherent multiple scattering. Thermodynamic disorder preserves the spin distribution from a single nanoparticle, and it significantly reduces the destructive interference and spin fluctuations that usually occur in coherent disorders. Because of this dynamical disorder, the spin field is constantly accumulated through many Brownian nanoparticles, leading to distinct spin fields homogeneously distributed at opposite sides of the laser (Fig. 1). Notably, the experimentally observed spin split can easily reach a centimeter scale, which is so-far the largest that has even been reported in disordered structures.

**Typical experimental results.** The experimental setting for observing the spin-locking effect is shown in Fig. 2A. The sample is hosted in a glass container, which is consisted of spherical gold nanoparticles (AuNPs) suspended in water (SI, section S7). The average diameter of AuNPs is $<D>$ = 250 nm (Fig. 2B). To observe the scattering effect, an *x*-polarized plane wave laser (wavelength

639 nm, beam width ~2 mm) impinges onto the sample at a normal incident angle. The observation branch is set perpendicular to the incident direction, and we use a simple lens system to capture the scattering images onto the camera. A quarter-wave plate and a post-polarizer are used to detect the intensities of opposite spin polarizations. All experiments are performed at a room temperature (~25°C), at which we testified the Brownian motion behavior of the samples using a DLS method (SI, section S11). As it shows in Fig. 2C, the second-order autocorrelation function $g_2$ indicates a strongly intensity fluctuation of the scattered light above a time scale of $\tau_c = 0.7$ ms. The acquisition time of our detector system is about 2 seconds; hence the observations are performed in the incoherent region. Typical examples of the observed intensity and spin distributions are shown in Fig. 2D and E, respectively. The scattering intensity distribution is composed of a ballistic region that spatially overlaps with the laser beam, and two diffusion regions at the opposite sides of the laser. The upper and lower diffusion regions show opposite spin angular momenta, respectively. Here, the spin (normalized) is characterized by $s_x = (I_R - I_L)/(I_R + I_L)$, with $I_{R,L}$ being the intensities for right- and left-handed circular polarizations. Despite the thermodynamic nature of the system, the observed spin split is very large and robust. Each spin region reaches a centimeter scale, and the spatial spin fluctuation in each region is very small (Fig. 3A, lower panel). Apparently, this unusual spin phenomenon is neither from a tilted optical illumination nor from the transverse spin of the laser (*26*), since we used a normally incident plane wave as excitation (Gaussian beam with a ~2mm beam waist). This spin effect potentially comes from the colloid. As a simple verification, we performed another experiment with the same optical setting but using a bottle of different AuNPs with $<D> = 400$ nm. As it shows in Fig. 2F and G, the sign of spin is flipped. Therefore, it indicates that the physical properties of the Brownian scattering system play a critical role in this spin-momentum locking effect. Additionally, this phenomenon is also demonstrated for nanoparticles of different material and concentration (SI, section S5 and S11).

**The incoherent multiple scattering theory**. To evaluate the observed BSLE from such a macroscale scattering system, we established a scattering theory that simplifies the complex scattering into two essential physical processes (SI, Fig. S15). (i) We calculate a two-step scattering process for a combination of any two nanoparticles: one located in the ballistic region ($\mathbf{r}_b$) and another in the diffusion region ($\mathbf{r}_d$). The light field of the incident laser, $\mathbf{E}_0$, excites one of the nanoparticles in the ballistic region, resulting in a radiation source $\mathbf{E}_b = \mathbf{M} \cdot \mathbf{E}_0$. Here, $\mathbf{M}$ stands for Mie scattering matrix, and $\mathbf{E}_b$ propagates in space following the strict solution (SI, section S8). In addition, another nanoparticle in the diffusion region $\mathbf{r}_d$ is similarly excited by $\mathbf{E}_b$, resulting in $\mathbf{E}_d = \mathbf{M} \cdot \mathbf{E}_b$. As a good approximation, we have $\mathbf{E}_d \propto \mathbf{E}_b(\mathbf{r}')$, with $\mathbf{r}' = \mathbf{r}_d - \mathbf{r}_b$. (ii) We calculate the field superposition of $\mathbf{E}_d$ from many nanoparticles at different $\mathbf{r}_b$. In the case of incoherent scattering, we project the field $\mathbf{E}_b(\mathbf{r}')$ onto the spin-up and spin-down states, and obtain $I_R(\mathbf{r}')$ and $I_L(\mathbf{r}')$, respectively. The intensities are then summed over all nanoparticles, $I_{R,L}(\mathbf{r}_d) = \sum_{\mathbf{r}_b} I_{R,L}(\mathbf{r}_d - \mathbf{r}_b)$. The macroscale spin effect is then shown by calculating the spatial distribution of the difference between $I_R(\mathbf{r}_d)$ and $I_L(\mathbf{r}_d)$.

In theory, we assumed a number of $N = 6 \times 10^5$ spherical AuNPs with the size of $D = 250$ nm, and these AuNPs are randomly assigned in a macroscale region (2 mm×2 mm×2 cm). Following the above-mentioned incoherent scattering theory, it results in intensity and spin distributions that show excellent agreements with our experiments (Fig. 3B with Fig. 2D and E). Moreover, the calculated spatial statistics in the diffusion regions are also matched with the experimental results (Fig. 3A). Particularly, the intensity probability distribution is a typical incoherent property, while the spin statistics are characterized by two narrowing Beta distributions with opposite central values. As a comparison, we also performed a coherent calculation using the same arrangement of AuNPs (SI, section S13). The results are drastically different from those of the incoherent calculation and experimental observations (Fig. 3C). Notably, many random speckles occur from the destructive interferences of coherent waves, with a significantly reduced spin-locking phenomenon. In general, the scattering theory can be developed to partial coherence to show the dynamical evolution of the system from incoherent to coherent scenarios. To do that, we divide a system of $N$ nanoparticles into $m$ parts. The intra-scattering within the same part is coherent, while the inter-scattering between different parts is incoherent. This simple mathematical approach provides us a parameter $m/N \in [0,1]$, through which we can calculate a clear evolution of the scattering property from incoherent (Fig. 3D, left) to coherent (Fig. 3D, right). It shows that strongest spin-locking phenomenon occurs in the incoherent scenario (with Brownian motion), while the increase of the degree of coherence will reduce the effect, and eventually eliminate the spin phenomenon from coherent disorder.

**Generic spin radiation of scattering**. The SOI of scattering originates from the non-diagonal components of **M**, indicating a universal effect when we convert a 2D-vector electromagnetic field into 3D via scattering. For extremely small particles in the Rayleigh limit, **M** is Green function operator. The scattered field is an electric dipole radiation, which exhibits purely transverse spin imbedded in the evanescent field, and $|\mathbf{s}| \propto 1/r^3$ (SI, section S8). The spin distribution of an electric dipole is composed of vortex-pair textures in radiation cones, similar to that for surface states of light and topological insulator (*13*) (SI, Fig. S8). This evanescent spin can be detected using nearfield coupling techniques (*27, 28*). In turn, dipole can be used as a probe to detect the evanescent spin from others as well (*29-31*). When the size of a nanoparticle increases, multipole response inevitably occurs. The scattered spin fields become radiative and partially aligned with the Poynting vector, and $|\mathbf{s}| \propto 1/r^2$ (SI, section S8). This effect is general: when a spherical nanoparticle is excited by a linearly polarized light, the radiation field is spin-dependent in an observation direction perpendicular to the incident wave, manifesting as an intrinsic geometric phase effect (SI, Eq. S6). We theoretically proved that this property emerges from out-of-phase couplings between any two harmonic modes of Mie scattering (SI, Eq. S4). Particularly, for metallic and dielectric nanoparticles, the electric quadrupole and magnetic dipole will arise to couple with the electric dipole, respectively. The latter scenario forms the well-known Janus dipole (*32*) (SI, section S8).

In Fig. 4A (the left-up panel), a typical example of the excited spin structure around an AuNP ($D$ = 250 nm) is shown. The vector $\mathbf{E}_0$ denotes the incident polarization of the laser. The spin vectors (normalized, $\mathbf{S}_\perp$) are spin angular momentum projected in the *xy* plane, and the colored circle shows the spin distribution in the radiation direction ($\mathbf{S}_\perp \cdot \hat{\rho}$). This four-lobe spin texture arises from the coupling between electric dipole, magnetic dipole and quadrupole, with the reddish and bluish colors meaning spin up and spin down states, respectively (SI, section S8). The dashed circular sectors mark the origin of the spin-dependent diffusion regions that are observed in our experiments (Fig. 4A, left-up panel, the experimental spin image). In particular, the size of the two circular sectors are determined by the numerical aperture of our lens system. The upper (lower) circular sector corresponds to the upper (lower) diffusion region captured by our camera (Fig. 4A). We testified the theory by continually changing the orientation of incident linear polarization, and we observe a good agreement between theory and experiments, as shown in Fig. 4A-C. Notably, due to the four-lobe spin distribution, if we switch the observation direction from +$x$ to –$x$, a spin flipped phenomenon can be observed (SI, section S1).

**Spin-resolved spectroscopy for nanoparticles' size measurement**. The spin-resolved optical field carries the information of individual nanoparticles by multiple scattering, which allows us to detect the physical properties of the Brownian nanoparticles from a paraxial optical imaging system. As a proof-of-demonstration, we developed a spin-resolved optical spectroscopy to measure the size of the AuNPs. To do that, we utilized a broadband laser as the excitation source, and the camera is replaced by a spectrometer, which captures $s_x(\lambda)$ (SI, section S12). We prepared three AuNPs samples with different nominated <$D$> (typical SEMs are shown in Fig. 5A), and we used SEM to measure their size distributions as a benchmark reference (the histograms in Fig. 5B). The experimentally observed spin-spectral $s_x(\lambda)$ are shown in Fig. 5C, in the wavelength range between 430 nm and 700 nm. The experimental $s_x(\lambda)$ is compared with different theoretical spectral, $s_{x,\text{theory}}(D, \lambda)$, which is calculated using the incoherent scattering theory applied for many AuNPs with size $D$. The normalized overlapping (correlation) between $s_x(\lambda)$ and $s_{x,\text{theory}}(D, \lambda)$ are depicted as the orange curves in Fig. 5B, and the stars mark the highest correlation for each sample. We see a good agreement between our method and the SEM. Particularly, for sample 1, the highest correlation occurs for $D$ = 100 nm, while that obtained from SEM is 101 nm. The accuracy from our spin-resolved spectroscopy is 99%. As for sample 2 and 3, the accuracy is 89% and 97%, respectively. The precision can be further improved by increasing the detection wavelength range, or using feedback algorithm for optimization.

**Discussion**

In summary, we have observed a spin-locking phenomenon of light from a Brownian system with spatiotemporal disorder. The system is consisted of enormous dynamical nanoparticles, serving as a bulk spin-probing medium that can transmit the spin field from single particles via many random scatterings. This phenomenon is universal and ubiquitous, and it can occur for nanoparticles with different shape, size or refractive index. The measurement of this spin effect can be developed as a spin-resolved optical means to detect the size, material, and other properties of nanoparticles using a simple paraxial optical imaging system. Since Brownian motions are widely existed, this method may find broad applications in chemistry, biology, micromechanics, or material science. Complex disordered systems offer a plethora of fascinating wave phenomena, and many of them are unexpected. Our study can be an inspiration for discovering similar effects in other wave-based systems for both classical and quantum regimes. One possible direction in the future is to investigate dynamical spin effects using fast optical detecting approaches (*33, 34*), or equivalently slowing down the Brownian motion via temperature or viscosity control. Moreover, the Brownian system can be formed of engineered nanostructures with customized size and shape. In this sense, the nanoparticles are thermally fluctuated meta-molecules, and a Brownian system is resembled as a liquid metamaterial to achieve exotic light-matter interactions.

**Author contributions**

E.H, B.W and X.C supervised this work. B.W initialized theory and experiment, observed the phenomena, and wrote the manuscript. X.Z systematically performed experimental work, theory, and figure preparation. P.C contributed significantly in Mie theory analysis, and assisted in experimental characterization. M.L performed $g_2$ and SEM measurement. Y.S assisted in manuscript revision. B. W, X. Z, P. C and M. L prepared the supplementary material. E.H and X.C evolved in discussions at all stages of this work and revised the manuscript. †X. Z and P. C contributed equally to this work.

**Acknowledgement**

We thank Xin Cao for helping us with sample preparation, and useful discussions with Yiming Pan at the early stage of this work. This work is supported by National Key Research and Development Program of China (2022YFA1205101), National Science Foundation of China (12274296, 12192252), Shanghai International Cooperation Program for Science and Technology (22520714300), Shanghai Jiao Tong University 2030 Initiative. B.W. is sponsored by Yangyang Development Fund. E.H. acknowledges financial support from the Israel Science Foundation (grant number 1170/20).

**Figures**

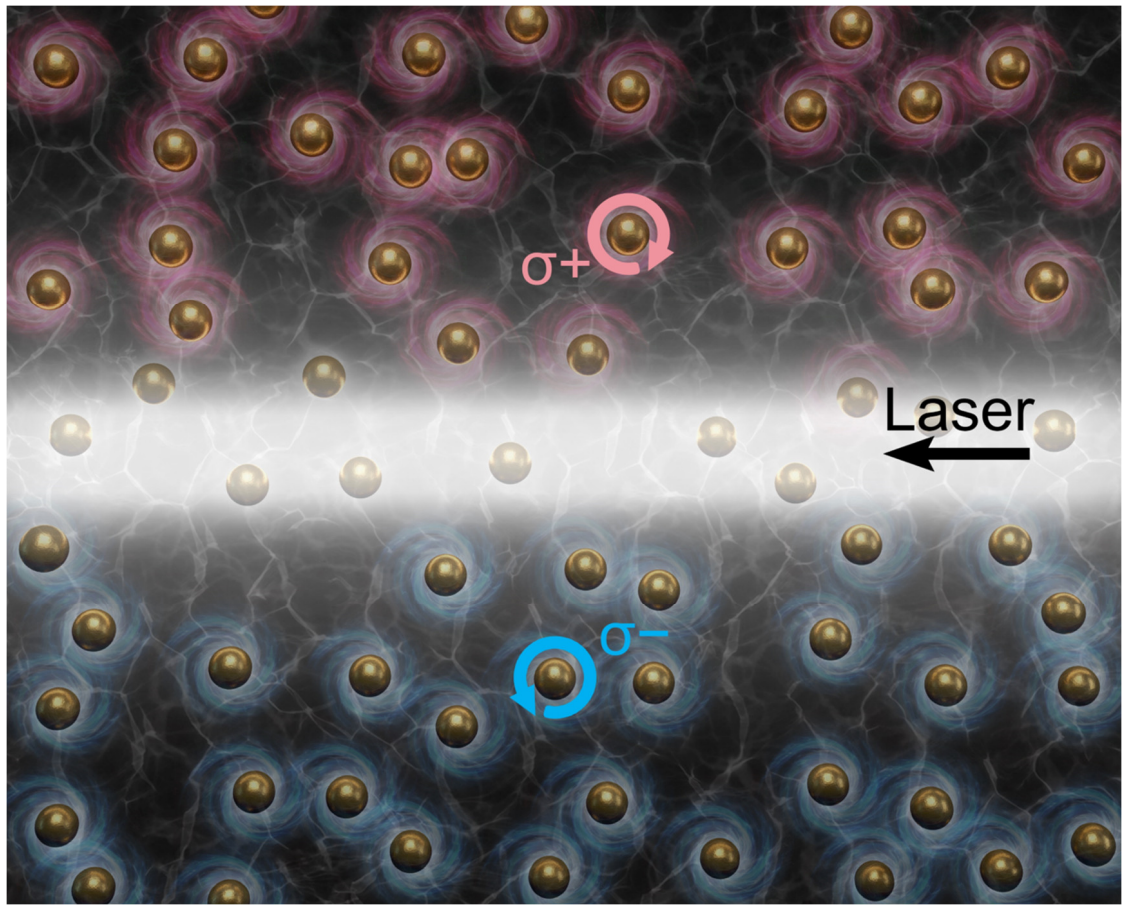

**Fig 1. Concept illustration of the BSLE.** A plane wave laser horizontally impinges onto a colloidal suspension from the right-hand side. The system is composed of many spherical gold nanoparticles in Brownian motion. The nanoparticles in the upper diffusive region all radiate a right-handed circular polarization, and the nanoparticles in the lower diffusive region all radiate a left-handed circular polarization.

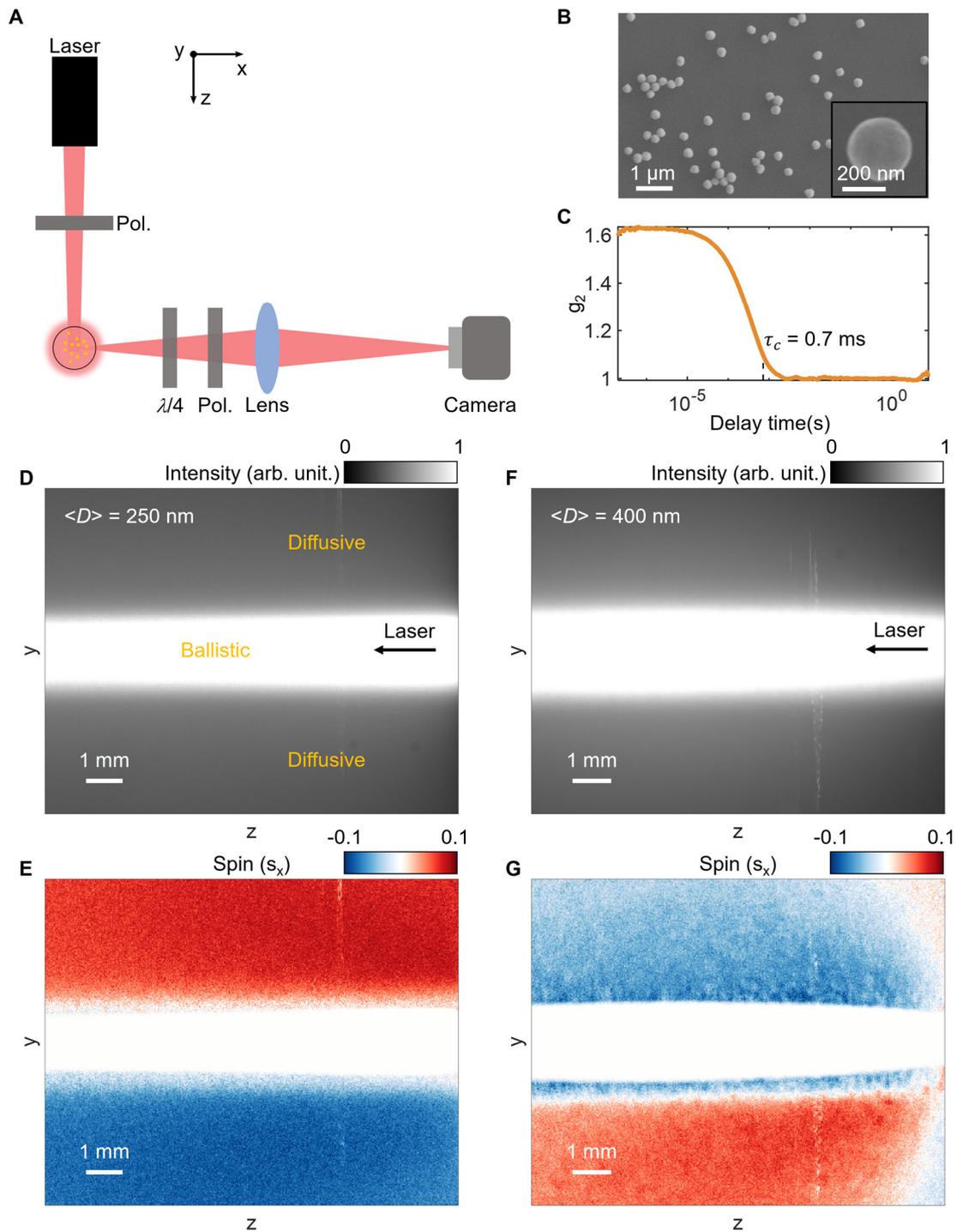

**Fig 2. Experimental observation of the BSLE of light scattered from Brownian nanoparticles.** (**A**) Experimental setup for observing the effect. A linearly polarized laser (639 nm) impinges onto the sample at a normal incident angle. The sample is a glass of AuNPs in water. The observation branch is set at a scattering angle perpendicular to the incident light. The image is captured by the camera from a single lens. Pol., linear polarizer; QWP, quarter-wave plate; Lens, a convex lens ($f$ = 80 mm). (**B**) An example of scanning electron microscope image of the AuNPs with $<D>$ = 250 nm. (**C**) Measured second-order autocorrelation function of the AuNPs using a DLS instrument. The coherence time is $\tau_c \approx 0.7$ ms. (**D** and **E**) Intensity and spin distributions of light scattering from of AuNPs ($<D>$ = 250 nm). (**F** and **G**) Intensity and spin distributions of light scattering from of AuNPs ($<D>$ = 400 nm).

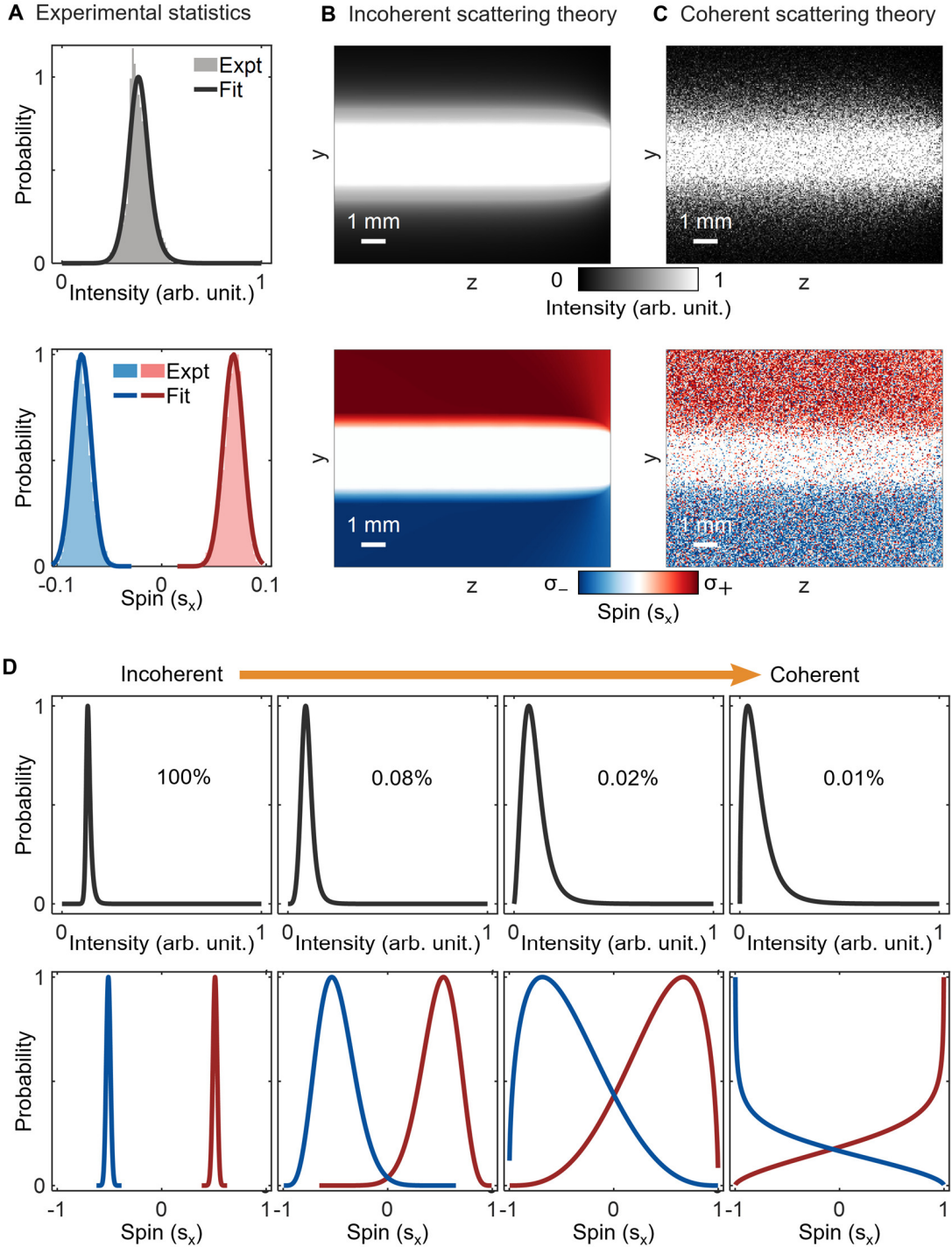

**Fig 3. Statistical properties of the BSLE and incoherent scattering theory.** (**A**) The histograms are experimentally observed statistical distributions of the spatial intensity (upper panel) and spin (lower panel), which are obtained from the diffusive regions of Fig. 2D and E, respectively. The solid curves are fitted by a Burr distribution for the intensity and Beta distributions for the spin. (**B**) The calculated spatial distributions of the normalized intensity and spin from the incoherent scattering theory. (**C**) The calculated spatial distributions of the normalized intensity and spin from the coherent scattering theory. (**D**) The theoretical evolution of the intensity and spin distributions by changing $m/N$ from 100% (incoherent) to 0.01% (coherent). As the degree of coherence increases, the spin distributions become wider with enhanced skewness, corresponding to increased spin fluctuations that reduces the spin-locking phenomenon.

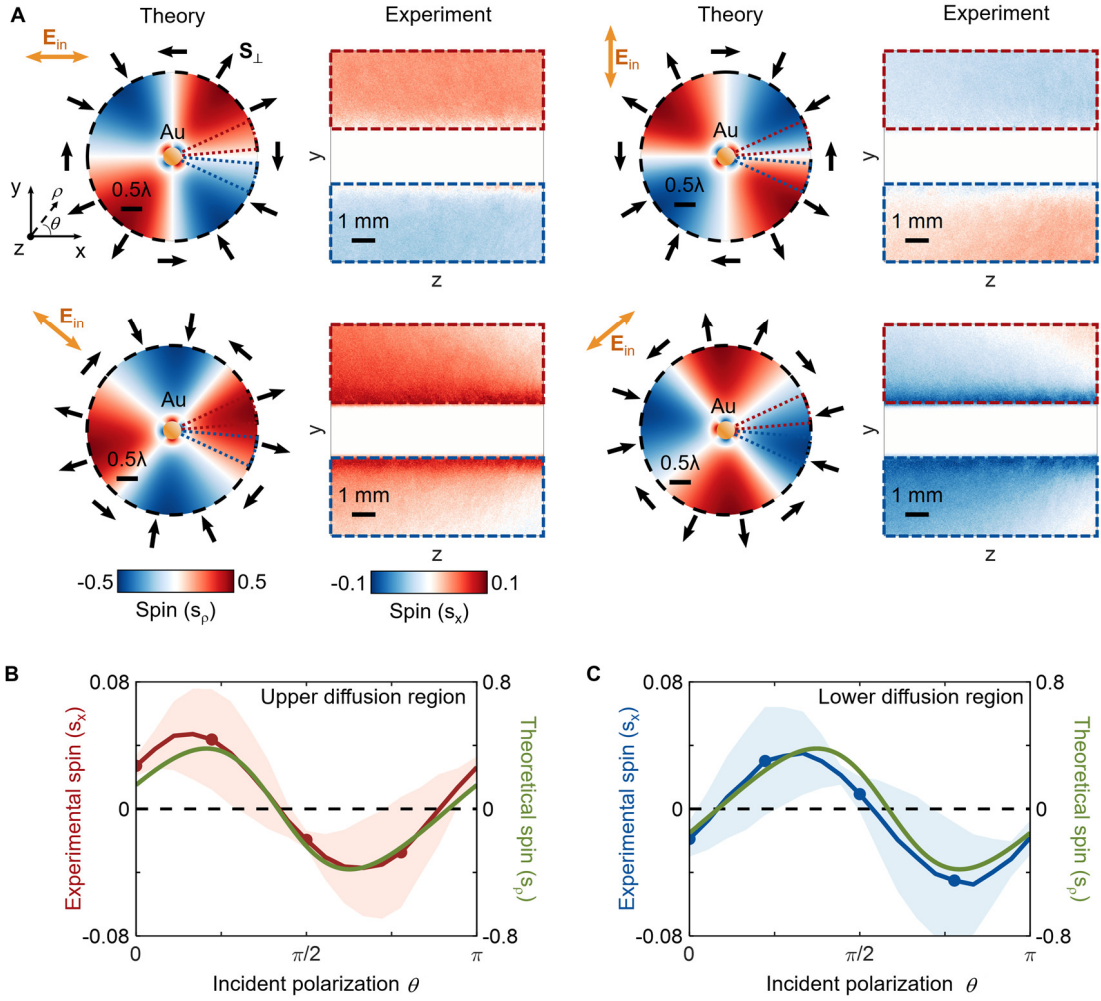

**Fig 4. The intrinsic spin from scattering and its polarization-dependent properties.** (A) The theoretical spin distribution of a single particle and the experimentally observed macroscale spin phenomena from different incident polarizations. The particle in theory is a gold nanosphere with $D = 250$ nm, which is excited by light at the wavelength of 639 nm. The double-headed orange arrows are the incident polarization orientations. The black arrows are the normalized spin components in the $xy$ plane. The blue and red colors represent the degree of spin and spin projected in $\hat{\rho}$. The two dashed circular sectors are the angular regions that correspond to the observed diffusion regions in our camera, which are determined by the numerical aperture of the lens system. In paraxial approximation ($\hat{\rho} \sim \hat{x}$), such as our imaging system, $s_\rho \propto s_x$. (**B** and **C**) Experimental (red and blue) and theoretical (green curves) results of the spin from different incident linear polarizations. The curves are the averaged results over the entire upper (B) or lower (C) diffusive region, and the shaded areas denote standard deviations, arising from the scattering in the $z$ direction.

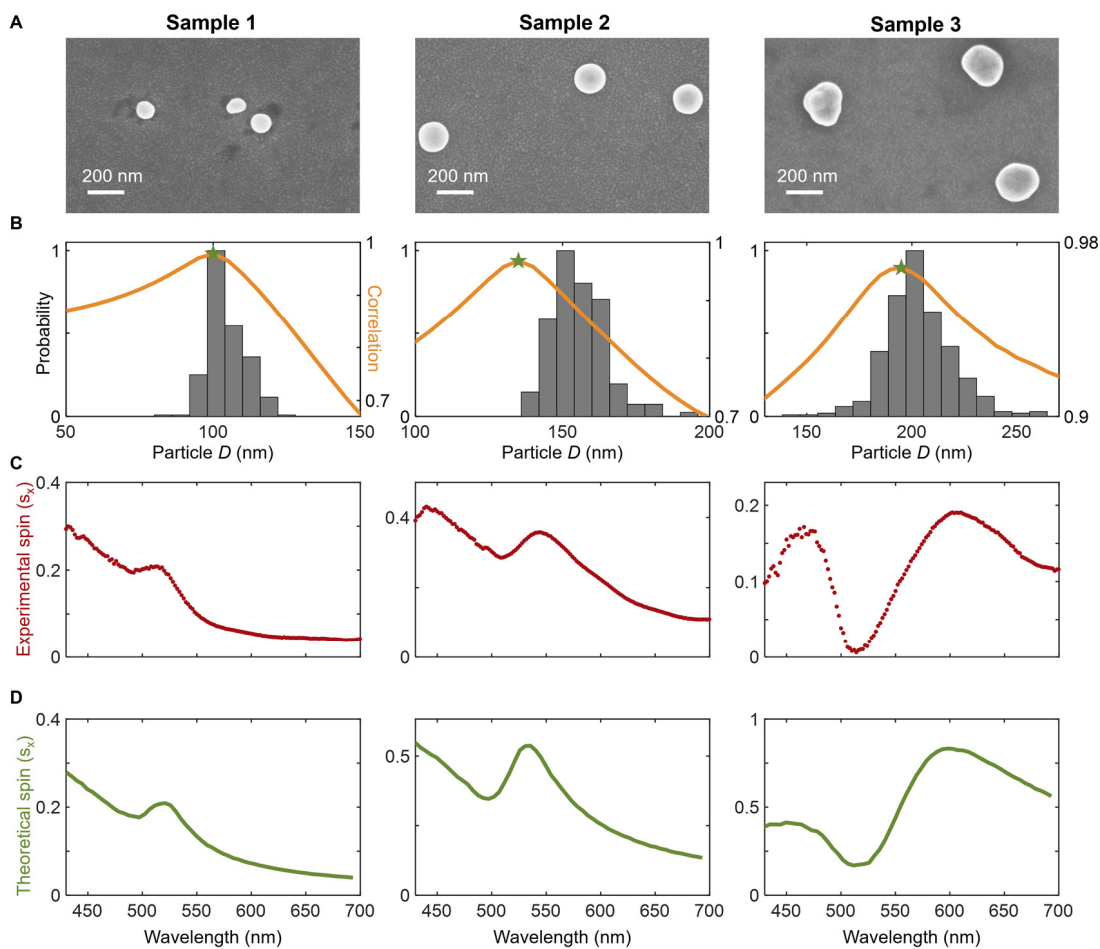

**Fig 5. Spin-resolved spectroscopy of BSLE for measuring the size of AuNPs.** (**A**) Typical examples of the SEM images from three different samples, with their nominated <*D*> = 100 nm, 150 nm and 200nm, respectively. (**B**) The histograms are measured *D* distributions of the three samples using SEM images. The curves are correlation functions that are defined by the overlapping between $s_{x,\text{theory}}(D, \lambda)$ and $s_x(\lambda)$, which is observed from the spin-resolved spectroscopy. The SEM-obtained highest probabilities regions are 98~104 nm, 148~154 nm and 197~205 nm, respectively. The highest correlations (marked as stars) obtained from the spin-resolved spectroscopy are 100 nm, 135 nm and 195 nm, respectively. (**C**) Experimentally observed $s_x(\lambda)$ from spin-resolved spectroscopy. (**D**) Examples of the theoretical spin-resolved spectra calculated using *D* = 100 nm, *D* = 135 nm and *D* = 195 nm, corresponding to the highest correlations in (B).

**Table of Contents**

# Section S1. Observation of the BSLE from opposite directions

As it shows in the main text figure 4A, the spin distribution from a single AuNP is a four-lobe pattern (Fig. S1). This is because the spin is defined as $\mathbf{S}_\perp$, the spin angular momentum projected onto the radial direction $\hat{\rho}$ in the *xy* plane. When the observation branch is along the +*x* direction (as we did in the main text), $s_\rho$ has the same sign with that of $\mathbf{s}_x$, which is defined by $(I_R - I_L)/(I_R + I_L)$ from +*x* direction. Here, $I_R$ and $I_L$ are the intensities of right- and left-handed circular polarization that are detected by inserting a quarter-wave plate and a linear polarizer between the sample and the camera. Note that this definition of handedness is associated with the propagation direction of light (momentum **k**), so that the spinning direction of the electric field for $I_R$ or $I_L$ is opposite for the +*x* and –*x* propagation light. This causes an interesting result depicted in Fig. S1: when we observe the scattering from an opposite direction, the spin distribution is flipped. The reddish color in the experimental images represents right-circular polarization, while the blueish color is for left-circular polarization (Fig. S1). The experimental demonstration is described below. We used a similar experimental setup as depicted in Fig. 2A. A linear polarizer is used between the laser ($\lambda$ = 639 nm) and the AuNPs sample to generate *x*-polarized incident light propagating along the *z* direction. We symmetrically set two observation branches to detect spin of scattered field on +*x* direction and –*x* direction. The lenses enable cameras to receive the images of AuNPs by focusing. The cameras captured a series of images at a frame rate of 100 ms and the exposure time is 30 ms.

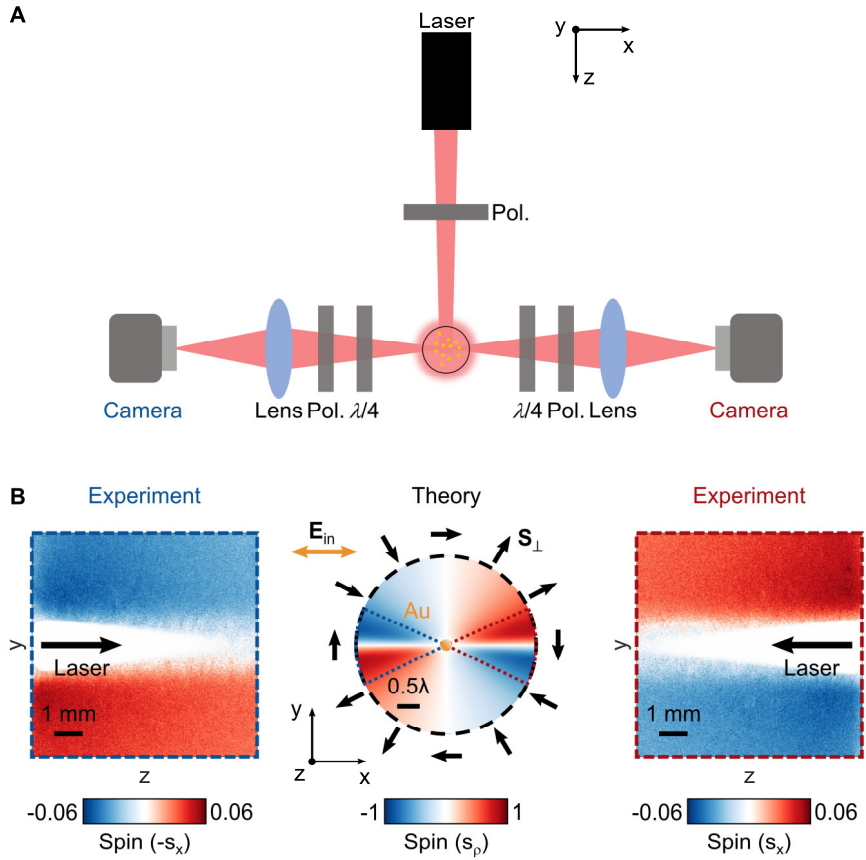

**Fig. S1 Experimental observation of the BSLE of light scattered from Brownian particles. (A)** Schematic of the optical setup. Laser, $\lambda = 639\ nm$; Pol., linear polarizer; $\lambda/4$, quarter-wave plate; Lens, focal length is 80 mm. **(B)** Middle: the four-lobe spin ($s_\rho$) distribution of the scattered field from a AuNP ($D = 150$ nm). The incident direction of laser is along +z direction. $\mathbf{E}_{in}$ represents the incident polarization. The double-headed black arrows are the normalized $\mathbf{S}_\perp$. Left and right: Experimentally observed spin-resolved images from –x direction (left) and +x direction (right). The reddish color stands for right-circular polarization, and the blueish color for left-circular polarization.

## Section S2. Comparison of BSLE from different beam waist and circular polarization

There are two key points to clarify in our BSLE. (1) This effect arises from the scattering of many nanoparticles, which is distinguished from the transverse spin effect of the incident laser beam. (2) It is worthy to demonstrate the behavior of the effect for circular polarization illumination. For these purposes, we conducted additional experiments in this section.

To show that our effect is independent of the transverse spin of the incident beam, we performed two groups of experiments, one is set in consistent with our main text setup using a plane wave (Gaussian beam waist, 2 mm), and the other uses a focused beam (focused beam waist < 0.5 mm). The experimental results are shown in Fig. S2A and B for two different polarization states. The results for plane wave and focused beam illumination are very similar, from both the spin distribution and the spin value. Therefore, the spin effect is not impacted by the beam waist of incident laser. We also provide a summarized theory for the transverse spin of focused beam in Section S9, where we can see many different properties from our BSLE compared to the transverse spin of focused laser. Additionally, we used opposite circular polarization as incident excitations to observe the phenomena (Fig. S2 C, D). In this case, the ballistic region shows a stronger spin value, while the diffusion region a weaker spin value. This weaker value effect is an indication of the depolarization process, which occurs because the scattering partly turns the incident right/left circular polarization to left/right circular polarization. We will present in Section S8D its relation with the geometric phase effect.

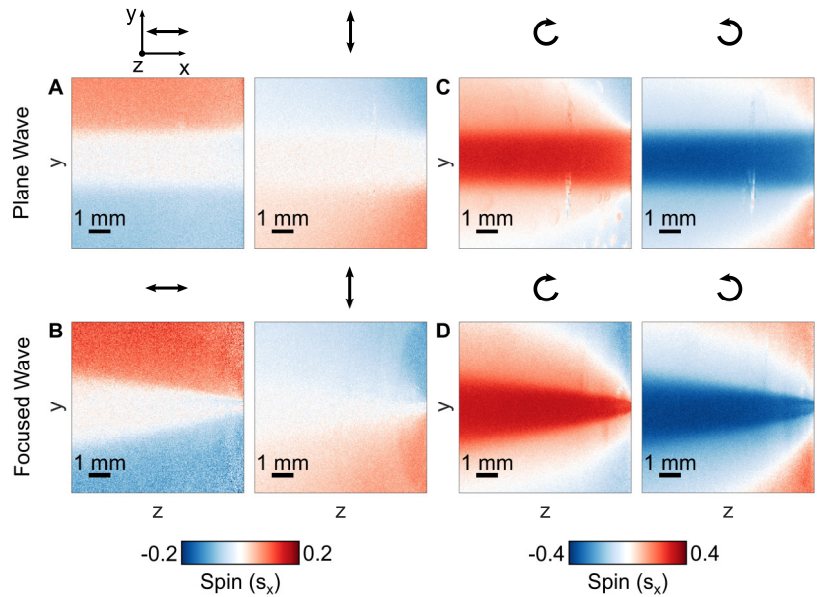

**Fig.S2 Observation from different beam waist and circular incident polarizations.** From left to right, *x* polarization, *y* polarization, right-handed and left-handed circular polarizations. The incident light is a plane wave in the first row, and a focused beam in the second row.

# Section S3. Experimental setup for the spin-resolved spectra measurement

Here we show the experimental details for the observation of main text Figure 5. We used a supercontinuum laser (YSL-photonics, SC-PRO-M) to illuminate the AuNPs and analyzed the spin resolved spectral at different diffusion regions (Fig. S3). The setup is illustrated in Fig. S3. A pinhole is placed onto the imaging plane to locate a region in the diffusive part for spectrum analysis. This location is ensured by the imaging captured by the camera in the second imaging plane. It should be noted that the spin distribution is almost homogeneous in each diffusion region, so it is easy to find a location away from the ballistic region for detection. After the pinhole is set, we used a spectrometer (Zolix Omni-$\lambda$5028i) to detect $I_R(\lambda)$ and $I_L(\lambda)$. Afterwards, $s_x(\lambda)$ is obtained from $[I_R(\lambda) - I_L(\lambda)]/[I_R(\lambda) + I_L(\lambda)]$.

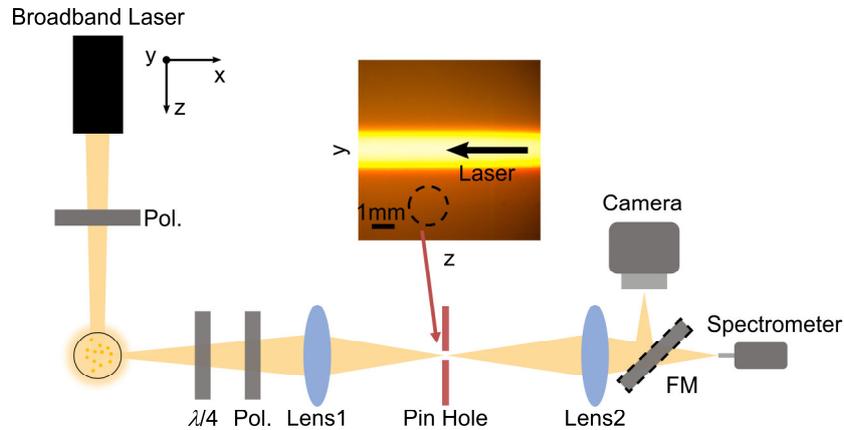

**Fig.S3 Schematic of experimental setup for spin-resolved spectroscopy.** We used a size-adjustable pinhole to select a small region (in the diffusion region) on the imaging plane to analyze the spin-resolved spectra (dashed circular region). The spectrometer (Zolix Omni-$\lambda$5028i) was used to capture the spin-resolved scattering spectrum of AuNPs. Broadband laser, *YSL*-photonics, SC-PRO-M; Pol., linear polarizer; $\lambda$/4, quarter-wave plate; FM, flip mirror; (Lens1, Lens2, 80 mm)

# Section S4. Observed BSLE from different linear polarizations

This section provides detailed experimental results for different incident polarizations,

as an extension of main text Figure 4. The spin images are shown in Fig. S4.

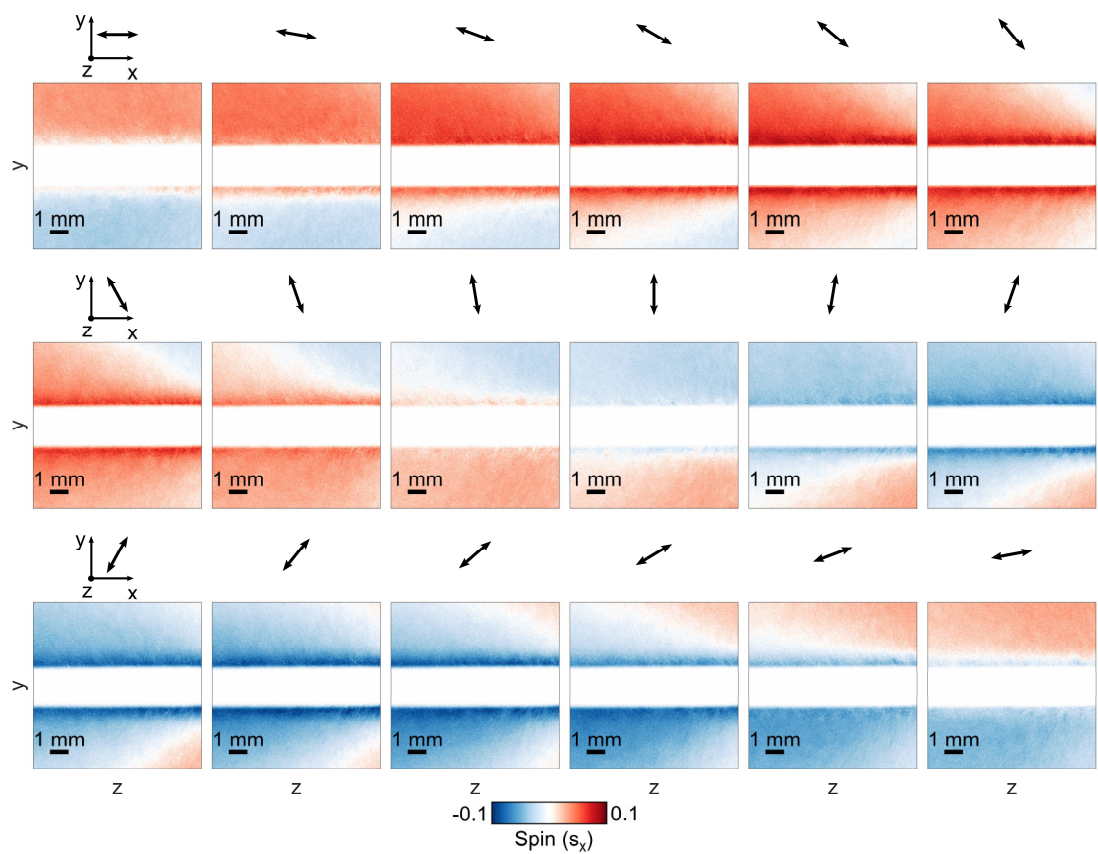

**Fig.S4 Detailed experimental results of the BSLE from different incident linear polarizations.** The black arrows on the top of each image represent the polarization of incident light in the *xy* plane. The images are captured in the *yz* plane, for the sample of AuNPs $<D>$ = 250 nm.

## Section S5. Observed BSLE from different concentrations of nanoparticles

Our main text scattering theory has simplified the multiple scattering process into two steps. This is essential for theoretical prediction of the BSLE from an enormous number of nanoparticle scattering. In reality, complex multiple scattering will occur to affect the results, which is described by a depolarization process (*1, 2*). Here, we experimentally show the behavior of BSLE from nanoparticles in different concentration (defined by the average number of nanoparticles in a unit volume). The experimental results show that even for high concentrations, the BSLE still survives, indicating its robustness against multiple dynamical scattering (Fig. S5). The details of experiments are presented below. The setup is the same to that in Fig. 2A, and the incident light is $x$-polarized. The tested samples are $Fe_3O_4$ nanoparticles (Fig. S13) with concentrations ranging from $0.57\times10^8$ cm$^{-3}$ to $5.70\times10^8$ cm$^{-3}$. As it shows in Fig. S5A, the ballistic region becomes weaker and eventually disappeared as the concentration increases. The scattering is completely diffusive within the sample, as the concentration reaches $5.70\times10^8$ cm$^{-3}$. This means that multiple scattering, rather than single or few scatterings, has dominated. The spin images are presented during this process, as shows in Fig. S5B. It can be seen that the BSLE is strong for small concentrations. As the concentration increases, the spin effect is weakened but the distribution is roughly remined. Specifically, for longer propagation distance, the spin decreases faster. We particularly take two points in the picture for analysis, ① and ②, which stands for a longer (~1 cm) and shorter (~1 mm) propagation distance in the scattering medium, and their intensity and spin evolution as a function of concentration are depicted in Fig. S5C and D, respectively. While the intensity evolution for both spots are very similar in this concentration region, the spin distribution is very different. Specifically, when the concentration of solution is approximately $2.28\times10^8$ cm$^{-3}$, the statistically averaged spin at ①, $\overline{|s_{x1}|}$, approaches 0, while $\overline{|s_{x2}|}$ is about 0.09. Notably, near the ② region, the observed $s_x$ distribution exists over a wide range of concentrations, indicating some robustness of the BLSE from single to multiple scattering process. Moreover, when we observe the ① region for a while, it is possible to see dynamical flip of the BSLE (Supplementary video).

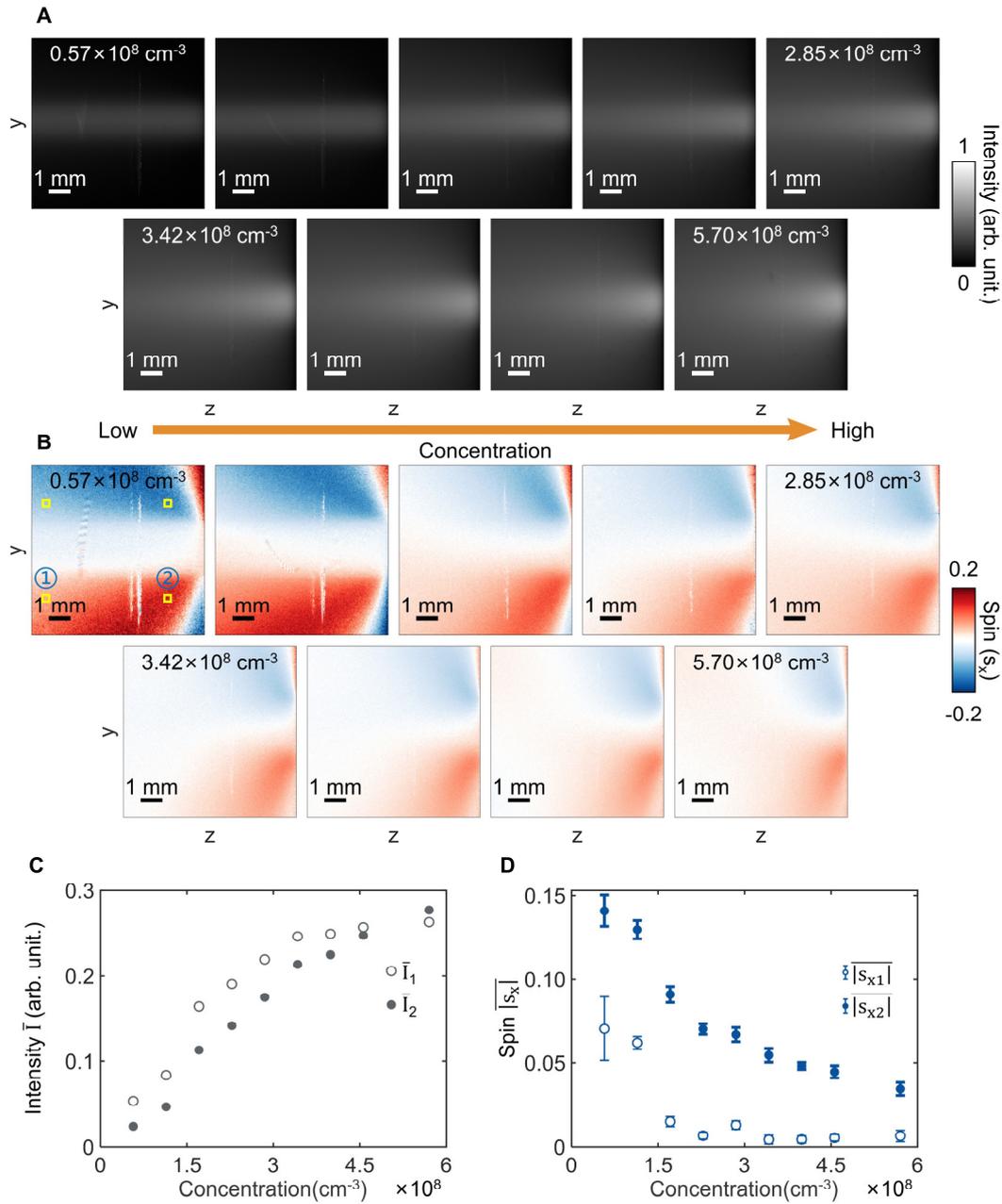

**Fig.S5 Observed BSLE from different concentration of $Fe_3O_4$ nanoparticles.** Normalized light intensity distribution **(A)** and $s_x$ distribution **(B)** from different concentration of $Fe_3O_4$ nanoparticles in water. The yellow boxes ①, ② represents the analysis region to obtain light intensity and $s_x$ in C and D. The upper two yellow boxes represent analysis regions that are symmetric with regions ①, ② with respect to the laser beam. Laser comes from the right side of these images. **(C)** The change of normalized light intensity with respect to concentration. $\bar{I}_1(\bar{I}_2)$ represent the average light intensity of region ① (②), respectively. **(D)** The change of $\overline{|s_x|}$ with respect to concentration. $\overline{|s_{x1}|}$ ($\overline{|s_{x2}|}$) represent the averaged absolute value of $s_x$ of region ① (②).

# Section S6. Extracting the value of spin from different detector exposure time

To confirm that the experimental value of $s_x$ is accurate, we have changed the exposure time of the camera and analyzed the evolution of $s_x$ to find a suitable exposure time and incident power. The experimental setup is same to that in Fig. 2A and the incident light is *x*-polarized. The power of incident laser is moderate, about 5 mW (continuous wave). The exposure time of the camera can be set from 0.1 ms to 12 ms. The intensities are shown in Fig. S6A, where we see the value increases with the exposure time. On the one hand, a sufficiently long exposure time allows us to observe a stable spin of the scattered light over a period within the Brownian motion nanoparticle system. On the other hand, we also need to avoid light intensity saturation, which also distorts the value of $s_x$. As shown in Fig. S6C and D, when the exposure time is longer than 3 ms, $\overline{|s_x|}$ is approximately stable in the diffusion region. This gives us a guideline to set the camera properly in order to obtain correct $s_x$. All experimental results are achieved after we perform this exposure time adjustment.

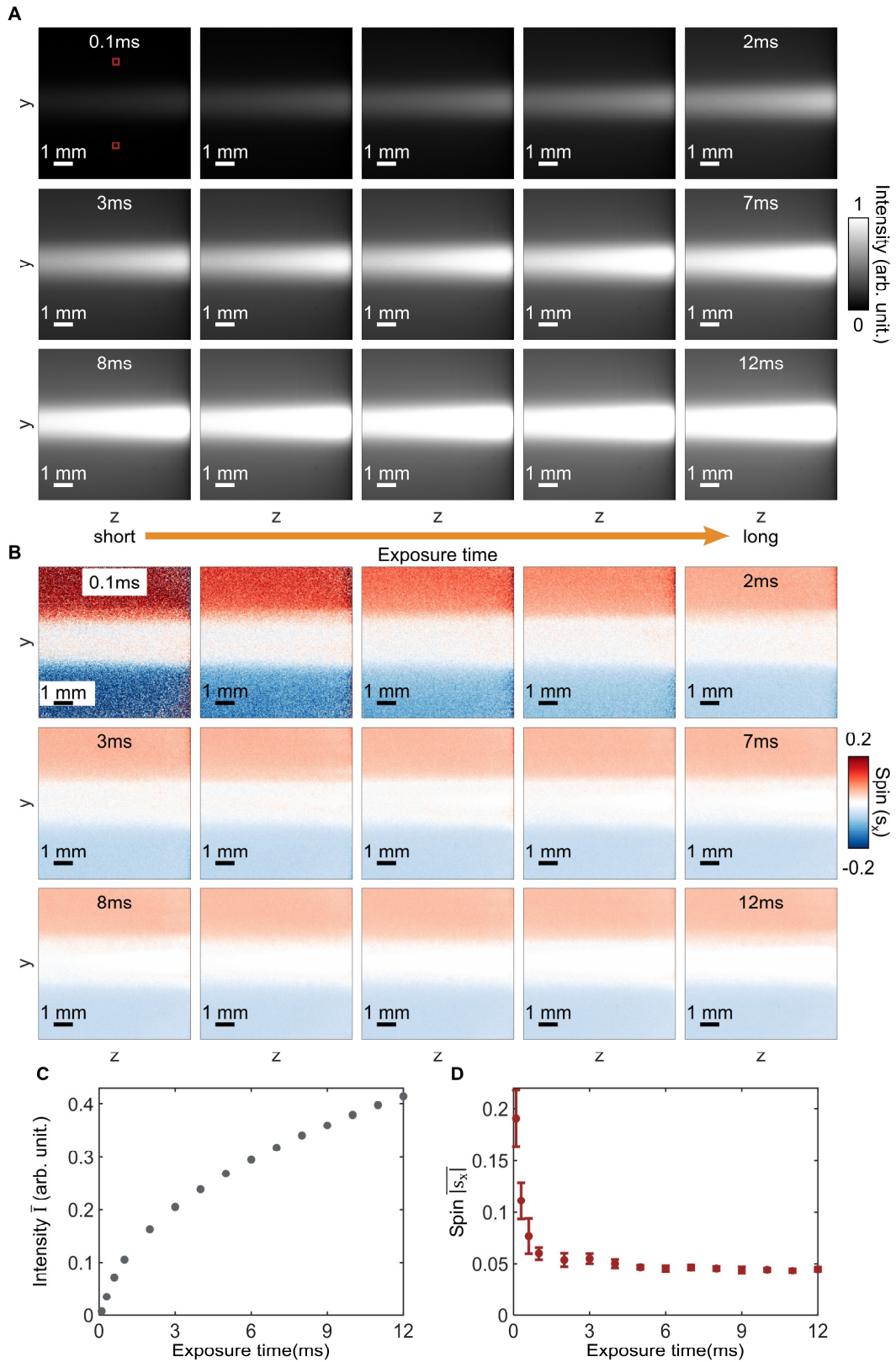

**Fig.S6 Captured $s_x$ and light intensity from different exposure time of camera.** Normalized light intensity distribution **(A)** and $s_x$ distribution **(B)** from different exposure time of camera. The red boxes in (A) represent the analysis regions to obtain light intensity and $s_x$ information for (C)(D). **(C)** The change of average light intensity

in the selected region with respect to the exposure time. **(D)** The change of $\overline{|s_x|}$ with respect to exposure time. $\overline{|s_x|}$ represents the average absolute value of $s_x$ of two red box regions.

## Section S7. The BSLE from different containers

The main experimental results were obtained by hosting the nanoparticles in a cylindrical glass container (Fig. S7). To show that the curved surface has negligible influence to our observed results, we also performed some experiments using cubic containers to host the samples, as shows in Fig. S7. The experimental setup is the same to it in Fig. 2A, and the incident light is $x$-polarized. The experimental results in Fig. S7 shows that the shape of container will not change the observation of our BSLE.

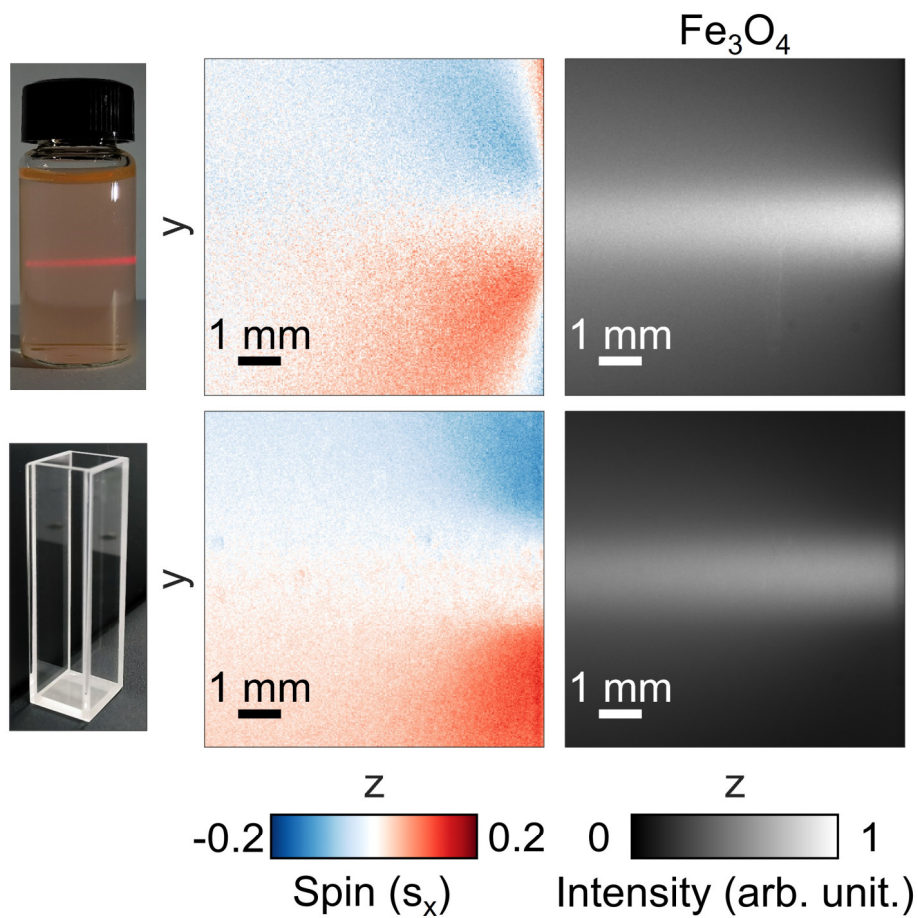

**Fig.S7 Experiments with different sample containers.**

# Section S8. Scattering theory

In this section, we present detailed theory derivations on the plane wave scattered from spherical nanoparticles (Mie theory). Particularly, we focus on the spin angular momentum distribution and spin-momentum locking relations. Several typical resonant modes are shown as examples. We will see the scattering system (3D electromagnetic field) has different spin properties as to those from 2D evanescent waves. This spin property also changes drastically from different harmonic modes or from their combinations. Therefore, multiple scattering provides a complicated environment for observing spin optical effects. In the following, we all assume the incident light is *x*-polarized plane wave propagating along *z*-axis.

## A. Spin-orbit relation of light for typical multipole scattering

Mie scattering describes the scattered field of a plane wave by a spherical particle (*3*). The solution to Maxwell equations can be expressed as a superposition of vector spherical harmonics (*3*), which corresponds to multipole expansion of electromagnetic field. The scattered electric and magnetic field are

$$\mathbf{E} = \sum_{n=1}^{\infty} E_n \left( i a_n \mathbf{N}_{e1n}^{(3)} - b_n \mathbf{M}_{o1n}^{(3)} \right),$$

$$\mathbf{H} = \frac{k}{\omega\mu} \sum_{n=1}^{\infty} E_n \left( i b_n \mathbf{N}_{o1n}^{(3)} + a_n \mathbf{M}_{e1n}^{(3)} \right). \qquad (S1)$$

Here, $E_n = \frac{i^n E_0 (2n+1)}{n(n+1)}$, $n \in \mathbf{Z}$. $E_0$ is the electric amplitude of the incident light, $k = \sqrt{\varepsilon\mu}\,\omega$ is the wave vector outside the particle, $\omega$ is the angular frequency, and $\varepsilon$ and $\mu$ represent the dielectric and magnetic permeabilities of the environment, respectively. The Mie coefficients $a_n$, $b_n$ represent the complex amplitude of electric and magnetic harmonics, respectively. They are determined by the incident wavelength, and the size and refractive index of the particle. The $\mathbf{N}_{e,o1n}^{(3)}$ and $\mathbf{M}_{e,o1n}^{(3)}$ are electric and magnetic vector spherical harmonics, respectively. The superscript (3) means that the radial part of the generating functions are spherical Hankel functions of the first kind (*3*). Conventionally, we discuss Mie scattering in a spherical coordinate $(r, \theta, \phi)$, where $r$ is the radial distance from the particle, $\theta$ is the polar angle and $\phi$ is the azimuthal angle. In the following subsections, we present several typical spin-orbit interactions of scattering from individual nanoparticles originated from the Mie scattering, derived from Eq. (S1).

## A1. Electric dipole

The simplest case of Mie scattering is an electric dipole approximation, also known as Rayleigh scattering. This happens when the size of a particle is very small compared to the incident wavelength. In this case, only $a_1$ is non-zero. The electric field $\mathbf{E}$, magnetic field $\mathbf{H}$, spin angular momentum (SAM) density $\mathbf{s} = \text{Im}\left[\frac{\varepsilon}{4\omega}\mathbf{E}^* \times \mathbf{E} + \frac{\mu}{4\omega}\mathbf{H}^* \times \mathbf{H}\right]$, Poynting vector $\mathbf{P} = \frac{1}{2}\text{Re}[\mathbf{E} \times \mathbf{H}^*]$, and the momentum density $\mathbf{p} = \varepsilon\mu\mathbf{P} = \frac{\varepsilon\mu}{2}\text{Re}[\mathbf{E} \times \mathbf{H}^*]$ can be derived from Eq. (S1), with the results concluded below,

$$\mathbf{E}_{a_1} = -E_1 a_1 e^{ikr}\left[\left(-\frac{2i}{k^2 r^2} + \frac{2}{k^3 r^3}\right)\cos\phi \sin\theta \,\hat{\mathbf{e}}_r - \left(-\frac{1}{kr} - \frac{i}{k^2 r^2} + \frac{1}{k^3 r^3}\right)\cos\phi \cos\theta \,\hat{\mathbf{e}}_\theta + \left(-\frac{1}{kr} - \frac{i}{k^2 r^2} + \frac{1}{k^3 r^3}\right)\sin\phi \,\hat{\mathbf{e}}_\phi\right],$$

$$\mathbf{H}_{a_1} = -\frac{k}{\omega\mu}E_1 a_1 e^{ikr}\left[\left(\frac{1}{kr} + \frac{i}{k^2 r^2}\right)\sin\phi \,\hat{\mathbf{e}}_\theta + \left(\frac{1}{kr} + \frac{i}{k^2 r^2}\right)\cos\phi \cos\theta \,\hat{\mathbf{e}}_\phi\right],$$

$$\mathbf{s}_{a_1} = \frac{\varepsilon}{2k\omega}|E_1|^2|a_1|^2\left[\frac{\sin 2\phi \sin\theta}{k^2 r^3}\hat{\mathbf{e}}_\theta + \frac{\cos^2\phi \sin 2\theta}{k^2 r^3}\hat{\mathbf{e}}_\phi\right],$$

$$\mathbf{P}_{a_1} = \frac{k}{2\omega\mu}|E_1|^2|a_1|^2\left[\frac{1}{k^2 r^2}(\cos^2\phi\cos^2\theta + \sin^2\phi)\hat{\mathbf{e}}_r\right],$$

$$\mathbf{p}_{a_1} = \frac{\varepsilon k}{2\omega}|E_1|^2|a_1|^2\left[\frac{1}{k^2 r^2}(\cos^2\phi\cos^2\theta + \sin^2\phi)\hat{\mathbf{e}}_r\right].$$

For an electric dipole, the SAM density of the scattered field is originated only from the electric field. This SAM density decays in space following a $1/r^3$ rule, hence it exists in the nearfield and dissipates away from the source. However, this $1/r^3$ decaying rule is still very different from the transverse spin of surface plasmon polaritons (or generally, surface evanescent waves), which decay exponentially. Particularly, we will find here the SAM density is locked to the momentum density with a unique relation:

$$\nabla \times \mathbf{p} = k^2 \mathbf{s}. \tag{S2}$$

One will notice the difference between Eq. (S2) and the established spin-momentum locking relation in reference (4), wherein $\nabla \times \mathbf{p} = 2k^2 \mathbf{s}$. This spin-momentum locking relation in Eq. (S2) is a unique property of dipole emitter in three-dimensional space. Previously, the study of spin-orbit interaction in scattering considered the dipole contribution, but without including the evanescent field (5).

### A2. Magnetic dipole

Magnetic dipole corresponds to the Mie coefficient $b_1$. While the incident light is $x$-polarized and propagating in the $z$-axis, the excited magnetic dipole is $y$-polarized. The electric field $\mathbf{E}$, magnetic field $\mathbf{H}$, SAM density $\mathbf{s}$, Poynting vector $\mathbf{P}$, and momentum density $\mathbf{p}$ of the scattered field are expressed below,

$$\mathbf{E}_{b_1} = -E_1 b_1 e^{ikr} \left[ \left( \frac{1}{kr} + \frac{i}{k^2 r^2} \right) \cos\phi \, \hat{\mathbf{e}}_\theta - \left( \frac{1}{kr} + \frac{i}{k^2 r^2} \right) \sin\phi \cos\theta \, \hat{\mathbf{e}}_\phi \right],$$

$$\mathbf{H}_{b_1} = -\frac{k}{\omega\mu} E_1 b_1 e^{ikr} \left[ \left( -\frac{2i}{k^2 r^2} + \frac{2}{k^3 r^3} \right) \sin\phi \sin\theta \, \hat{\mathbf{e}}_r - \left( -\frac{1}{kr} - \frac{i}{k^2 r^2} + \frac{1}{k^3 r^3} \right) \sin\phi \cos\theta \, \hat{\mathbf{e}}_\theta - \left( -\frac{1}{kr} - \frac{i}{k^2 r^2} + \frac{1}{k^3 r^3} \right) \cos\phi \, \hat{\mathbf{e}}_\phi \right],$$

$$\mathbf{s}_{b_1} = \frac{\varepsilon}{2k\omega} |E_1|^2 |b_1|^2 \left[ -\frac{\sin 2\phi \sin\theta}{k^2 r^3} \hat{\mathbf{e}}_\theta + \frac{\sin^2\phi \sin 2\theta}{k^2 r^3} \hat{\mathbf{e}}_\phi \right],$$

$$\mathbf{P}_{b_1} = \frac{k}{2\omega\mu} |E_1|^2 |b_1|^2 \left[ \frac{1}{k^2 r^2} (\cos^2\theta \sin^2\phi + \cos^2\phi) \hat{\mathbf{e}}_r \right],$$

$$\mathbf{p}_{b_1} = \frac{\varepsilon k}{2\omega} |E_1|^2 |b_1|^2 \left[ \frac{1}{k^2 r^2} (\cos^2\theta \sin^2\phi + \cos^2\phi) \hat{\mathbf{e}}_r \right].$$

The results are similar to those obtained from the electric dipole because of the electric-magnetic symmetry. For the magnetic dipole, the SAM density of the scattered field originates only from the magnetic field. Similarly, the SAM density is perpendicular to momentum density and they satisfy the same spin-momentum locking relation described by Eq. (S2).

### A3. Electric quadrupole

The electric quadrupole corresponds to Mie coefficient $a_2$. The electric field **E**, magnetic field **H**, SAM density **s**, Poynting vector **P** and momentum density **p** of scattered field are expressed as

$$\mathbf{E}_{a_2} = E_2 a_2 e^{ikr} \left[ -9\left(-\frac{1}{k^2r^2} - \frac{3i}{k^3r^3} + \frac{3}{k^4r^4}\right) \cos\phi \sin 2\theta\, \hat{\mathbf{e}}_r - 3\left(-\frac{i}{kr} + \frac{3}{k^2r^2} + \frac{6i}{k^3r^3} - \frac{6}{k^4r^4}\right) \cos\phi \cos 2\theta\, \hat{\mathbf{e}}_\theta + 3\left(-\frac{i}{kr} + \frac{3}{k^2r^2} + \frac{6i}{k^3r^3} - \frac{6}{k^4r^4}\right) \sin\phi \cos\theta\, \hat{\mathbf{e}}_\phi \right],$$

$$\mathbf{H}_{a_2} = \frac{k}{\omega\mu} E_2 a_2 e^{ikr} \left[ 3\left(\frac{i}{kr} - \frac{3}{k^2r^2} - \frac{3i}{k^3r^3}\right) \sin\phi \cos\theta\, \hat{\mathbf{e}}_\theta + 3\left(\frac{i}{kr} - \frac{3}{k^2r^2} - \frac{3i}{k^3r^3}\right) \cos\phi \cos 2\theta\, \hat{\mathbf{e}}_\phi \right],$$

$$\mathbf{s}_{a_2} = \frac{\varepsilon}{2k\omega} |E_2|^2 |a_2|^2 \left[ \frac{27 \sin 2\phi \sin 2\theta \cos\theta}{2k^2r^3} \hat{\mathbf{e}}_\theta + \frac{27\cos^2\phi \sin 4\theta}{2k^2r^3} \hat{\mathbf{e}}_\phi \right],$$

$$\mathbf{P}_{a_2} = \frac{k}{2\omega\mu} |E_2|^2 |b_2|^2 \left[ \frac{9}{k^2r^2} (\cos^2 2\theta \cos^2\phi + \cos^2\theta \sin^2\phi) \hat{\mathbf{e}}_r \right],$$

$$\mathbf{p}_{a_2} = \frac{\varepsilon k}{2\omega} |E_2|^2 |b_2|^2 \left[ \frac{9}{k^2r^2} (\cos^2 2\theta \cos^2\phi + \cos^2\theta \sin^2\phi) \hat{\mathbf{e}}_r \right].$$

In this case, the SAM density is still perpendicular to the momentum density. However, they are not constrained by the spin-momentum locking relation of Eq. (S2).

### A4. Magnetic quadrupole

The magnetic quadrupole is from the Mie coefficient $b_2$. The electric field **E**, magnetic field **H**, SAM density **s**, Poynting vector **P** and momentum density **p** of scattered field are expressed as below,

$$\mathbf{E}_{b_2} = E_2 b_2 e^{ikr} \left[ 3\left(\frac{i}{kr} - \frac{3}{k^2r^2} - \frac{3i}{k^3r^3}\right) \cos\phi \cos\theta\, \hat{\mathbf{e}}_\theta - 3\left(\frac{i}{kr} - \frac{3}{k^2r^2} - \frac{3i}{k^3r^3}\right) \sin\phi \cos 2\theta\, \hat{\mathbf{e}}_\phi \right],$$

$$\mathbf{H}_{b_2} = \frac{k}{\omega\mu} E_2 b_2 e^{ikr} \left[ -9\left(-\frac{1}{k^2r^2} - \frac{3i}{k^3r^3} + \frac{3}{k^4r^4}\right) \sin\phi \sin 2\theta\, \hat{\mathbf{e}}_r - 3\left(-\frac{i}{kr} + \frac{3}{k^2r^2} + \right.\right.$$

$$\left.\frac{6i}{k^3r^3} - \frac{6}{k^4r^4}\right)\sin\phi\cos 2\theta\,\hat{e}_\theta - 3\left(-\frac{i}{kr} + \frac{3}{k^2r^2} + \frac{6i}{k^3r^3} - \frac{6}{k^4r^4}\right)\cos\phi\cos\theta\,\hat{e}_\phi\Bigg],$$

$$\mathbf{s}_{b_2} = \frac{\varepsilon}{2k\omega}|E_2|^2|b_2|^2\left[-\frac{27\sin 2\phi\sin 2\theta\cos\theta}{2k^2r^3}\hat{e}_\theta + \frac{27\sin^2\phi\sin 4\theta}{2k^2r^3}\hat{e}_\phi\right],$$

$$\mathbf{P}_{b_2} = \frac{k}{2\omega\mu}|E_2|^2|b_2|^2\left[\frac{9}{k^2r^2}(\cos^2 2\theta\sin^2\phi + \cos^2\theta\cos^2\phi)\hat{e}_r\right],$$

$$\mathbf{p}_{b_2} = \frac{\varepsilon k}{2\omega}|E_2|^2|b_2|^2\left[\frac{9}{k^2r^2}(\cos^2 2\theta\sin^2\phi + \cos^2\theta\cos^2\phi)\hat{e}_r\right].$$

Akin to the electric quadrupole, the SAM density here is still always perpendicular to the momentum density. However, they are not quantitatively constrained by the spin-momentum locking relation of Eq. (S2).

### A5. Huygens dipole

While a single mode analysis is theoretically interesting for the understanding of the basic spin properties of scattering, the interaction between different mode is more common for natural particles. Here, we consider multipole response of the particle with Mie coefficient $a_1 = b_1$. This emitter is known as a Huygens dipole (6), which meets a Kerker condition (7). The electric field $\mathbf{E}$, magnetic field $\mathbf{H}$, SAM density $\mathbf{s}$, Poynting vector $\mathbf{P}$ and momentum density $\mathbf{p}$ of scattered field are expressed as below,

$$\mathbf{E} = \mathbf{E}_{a_1} + \mathbf{E}_{b_1} = -E_1 a_1 e^{ikr}\Bigg\{\left(-\frac{2i}{k^2r^2} + \frac{2}{k^3r^3}\right)\cos\phi\sin\theta\,\hat{e}_r + \left[\left(\frac{1}{kr} + \frac{i}{k^2r^2} - \frac{1}{k^3r^3}\right)\cos\phi\cos\theta + \left(\frac{1}{kr} + \frac{i}{k^2r^2}\right)\cos\phi\right]\hat{e}_\theta - \left[\left(\frac{1}{kr} + \frac{i}{k^2r^2} - \frac{1}{k^3r^3}\right)\sin\phi + \left(\frac{1}{kr} + \frac{i}{k^2r^2}\right)\sin\phi\cos\theta\right]\hat{e}_\phi\Bigg\},$$

$$\mathbf{H} = \mathbf{H}_{a_1} + \mathbf{H}_{b_1} = -\frac{k}{\omega\mu}E_1 a_1 e^{ikr}\Bigg\{\left(-\frac{2i}{k^2r^2} + \frac{2}{k^3r^3}\right)\sin\phi\sin\theta\,\hat{e}_r + \left[\left(\frac{1}{kr} + \frac{i}{k^2r^2} - \frac{1}{k^3r^3}\right)\sin\phi\cos\theta + \left(\frac{1}{kr} + \frac{i}{k^2r^2}\right)\sin\phi\right]\hat{e}_\theta + \left[\left(\frac{1}{kr} + \frac{i}{k^2r^2} - \frac{1}{k^3r^3}\right)\cos\phi + \left(\frac{1}{kr} + \frac{i}{k^2r^2}\right)\cos\phi\cos\theta\right]\hat{e}_\phi\Bigg\},$$

$$\mathbf{s} = \mathbf{s}_{a_1} + \mathbf{s}_{b_1} + \mathbf{s}_{coupling}$$

$$\mathbf{s}_{a_1} + \mathbf{s}_{b_1} = \frac{\varepsilon}{2k\omega} |E_1|^2 |a_1|^2 \frac{\sin 2\theta}{k^2 r^3} \hat{\mathbf{e}}_\phi,$$

$$\mathbf{s}_{coupling} = \frac{\varepsilon}{2k\omega} |E_1|^2 |a_1|^2 \left( \frac{2\sin\theta}{k^2 r^3} + \frac{2\sin\theta}{k^4 r^5} \right) \hat{\mathbf{e}}_\varphi,$$

$$\mathbf{s} = \frac{\varepsilon}{2k\omega} E_0^2 |a_1|^2 \left[ \frac{9}{4} \left( \frac{2\sin\theta(1+\cos\theta)}{k^2 r^3} + \frac{2\sin\theta}{k^4 r^5} \right) \hat{\mathbf{e}}_\phi \right],$$

$$\mathbf{P} = \mathbf{P}_{a_1} + \mathbf{P}_{b_1} + \mathbf{P}_{coupling}$$

$$\mathbf{P}_{a_1} + \mathbf{P}_{b_1} = \frac{k}{2\omega\mu} |E_1|^2 |a_1|^2 \frac{(\cos^2\theta + 1)}{k^2 r^2} \hat{\mathbf{e}}_r,$$

$$\mathbf{P}_{coupling} = \frac{k}{2\omega\mu} |E_1|^2 |a_1|^2 \left[ \left( \frac{2\cos\theta}{k^2 r^2} + \frac{\cos\theta}{k^6 r^6} \right) \hat{\mathbf{e}}_r + \frac{2\sin\theta}{k^6 r^6} \hat{\mathbf{e}}_\theta \right],$$

$$\mathbf{P} = \frac{k}{2\omega\mu} E_0^2 |a_1|^2 \frac{9}{4} \left\{ \left[ \frac{(1+\cos\theta)^2}{k^2 r^2} + \frac{\cos\theta}{k^6 r^6} \right] \hat{\mathbf{e}}_r + \frac{2\sin\theta}{k^6 r^6} \hat{\mathbf{e}}_\theta \right\},$$

$$\mathbf{p} = \frac{\varepsilon k}{2\omega} E_0^2 |a_1|^2 \frac{9}{4} \left\{ \left[ \frac{1}{k^2 r^2} (1+\cos\theta)^2 + \frac{1}{k^6 r^6} \cos\theta \right] \hat{\mathbf{e}}_r + \frac{2\sin\theta}{k^6 r^6} \hat{\mathbf{e}}_\theta \right\}.$$

The SAM and momentum are a mixture of individual contributions from the electric dipole, magnetic dipole, and their interference (coupling). Through the entire space, the individual contributions from $a_1$ and $b_1$ satisfy the spin-momentum locking relation: $\nabla \times (\mathbf{p}_{a_1} + \mathbf{p}_{b_1}) = k^2 (\mathbf{s}_{a_1} + \mathbf{s}_{b_1})$. For the spatial region with negligible $1/r^5$ and $1/r^6$, we have $\nabla \times \mathbf{p}_{coupling} = \nabla \times (\varepsilon \mu \mathbf{P}_{coupling}) = k^2 \mathbf{s}_{coupling}$. Therefore, the spin-momentum locking relation **Eq. (S2)** also works for Huygens dipole, as long as we can ignore the $1/r^5$ and $1/r^6$ terms.

### A6. Janus dipole

Aside from the Huygens dipole, there is another very interesting scattering known as Janus dipole, which is defined by $b_1 = ia_1$. The Janus dipole has been studied for unidirectional spin coupling and unusual nearfield spin/momentum properties (*8-12*). Here, we give the strict electric field **E**, magnetic field **H**, SAM density **s**, Poynting vector **P**, and momentum density **p** of the Janus dipole,

$$\mathbf{E} = \mathbf{E}_{a_1} + \mathbf{E}_{b_1} = -E_1 a_1 e^{ikr} \left\{ \left( -\frac{2i}{k^2 r^2} + \frac{2}{k^3 r^3} \right) \cos\phi \sin\theta\, \hat{\mathbf{e}}_r + \left[ \left( \frac{1}{kr} + \frac{i}{k^2 r^2} - \frac{1}{k^3 r^3} \right) \cos\phi \cos\theta + i \left( \frac{1}{kr} + \frac{i}{k^2 r^2} \right) \cos\phi \right] \hat{\mathbf{e}}_\theta - \left[ \left( \frac{1}{kr} + \frac{i}{k^2 r^2} - \frac{1}{k^3 r^3} \right) \sin\phi + i \left( \frac{1}{kr} + \frac{i}{k^2 r^2} \right) \sin\phi \cos\theta \right] \hat{\mathbf{e}}_\phi \right\},$$

$$\mathbf{H} = \mathbf{H}_{a_1} + \mathbf{H}_{b_1} = -\frac{k}{\omega\mu} E_1 a_1 e^{ikr} \left\{ i\left( -\frac{2i}{k^2 r^2} + \frac{2}{k^3 r^3} \right) \sin\phi \sin\theta\, \hat{\mathbf{e}}_r + \left[ i\left( \frac{1}{kr} + \frac{i}{k^2 r^2} - \frac{1}{k^3 r^3} \right) \sin\phi \cos\theta + \left( \frac{1}{kr} + \frac{i}{k^2 r^2} \right) \sin\phi \right] \hat{\mathbf{e}}_\theta + \left[ i\left( \frac{1}{kr} + \frac{i}{k^2 r^2} - \frac{1}{k^3 r^3} \right) \cos\phi + \left( \frac{1}{kr} + \frac{i}{k^2 r^2} \right) \cos\phi \cos\theta \right] \hat{\mathbf{e}}_\phi \right\},$$

$$\mathbf{s} = \mathbf{s}_{a_1} + \mathbf{s}_{b_1} + \mathbf{s}_{coupling}$$

$$\mathbf{s}_{a_1} + \mathbf{s}_{b_1} = \frac{\varepsilon}{2k\omega} |E_1|^2 |a_1|^2 \frac{\sin 2\theta}{k^2 r^3} \hat{\mathbf{e}}_\phi,$$

$$\mathbf{s}_{coupling} = \frac{\varepsilon}{2k\omega} |E_1|^2 |a_1|^2 \frac{\sin^2\theta \sin 2\phi}{kr^2} \hat{\mathbf{e}}_r,$$

$$\mathbf{s} = \frac{\varepsilon}{2k\omega} E_0^2 |a_1|^2 \left[ \frac{9}{4} \left( \frac{\sin^2\theta \sin 2\phi}{kr^2} \hat{\mathbf{e}}_r + \frac{\sin 2\theta}{k^2 r^3} \hat{\mathbf{e}}_\phi \right) \right],$$

$$\mathbf{P} = \mathbf{P}_{a_1} + \mathbf{P}_{b_1} + \mathbf{P}_{coupling}$$

$$\mathbf{P}_{a_1} + \mathbf{P}_{b_1} = \frac{k}{2\omega\mu} |E_1|^2 |a_1|^2 \frac{(\cos^2\theta + 1)}{k^2 r^2} \hat{\mathbf{e}}_r,$$

$$\mathbf{P}_{coupling} = \frac{k}{2\omega\mu} |E_1|^2 |a_1|^2 \left[ \frac{2 \cos 2\phi \sin\theta}{k^2 r^2} \hat{\mathbf{e}}_\theta - \frac{\sin 2\phi \sin 2\theta}{k^3 r^3} \hat{\mathbf{e}}_\phi \right],$$

$$\mathbf{P} = \frac{k}{2\omega\mu} E_0^2 |a_1|^2 \left[ \frac{9}{4} \left( \frac{1+\cos^2\theta}{k^2 r^2} \hat{\mathbf{e}}_r + \frac{2 \cos 2\phi \sin\theta}{k^3 r^3} \hat{\mathbf{e}}_\theta - \frac{\sin 2\phi \sin 2\theta}{k^3 r^3} \hat{\mathbf{e}}_\phi \right) \right],$$

$$\mathbf{p} = \frac{\varepsilon k}{2\omega} E_0^2 |a_1|^2 \left[ \frac{9}{4} \left( \frac{1+\cos^2\theta}{k^2 r^2} \hat{\mathbf{e}}_r + \frac{2 \cos 2\phi \sin\theta}{k^3 r^3} \hat{\mathbf{e}}_\theta - \frac{\sin 2\phi \sin 2\theta}{k^3 r^3} \hat{\mathbf{e}}_\phi \right) \right].$$

We can see that the SAM of a Janus dipole is a far field effect ($\mathbf{s} \sim 1/r^2$), which means it can be detected even if $r \gg \lambda$. The uncoupled terms of SAM density and the momentum density satisfy the spin-orbit relation: $\nabla \times (\mathbf{p}_{a_1} + \mathbf{p}_{b_1}) = k^2 (\mathbf{s}_{a_1} + \mathbf{s}_{b_1})$. Additionally, SAM is in parallel or antiparallel to the momentum in the far field, depending on the sign of the spatial coordinate $\phi$. Therefore, we have

$$\mathbf{s} = \frac{9\varepsilon}{8k\omega} E_0^2 |a_1|^2 \left( \frac{\sin^2\theta}{kr^2} \hat{\mathbf{e}}_r \right) \sin 2\phi. \tag{S3}$$

From Eq. (S3), we can predict a momentum-space spin Hall effect of light from a Janus dipole scattering, while the spin is depending on the observation angle of $\phi$.

### A7. In-phase combination of electric and magnetic quadrupole

For Mie coefficient $a_2 = b_2$, the electric field $\mathbf{E}$, magnetic field $\mathbf{H}$, SAM density $\mathbf{s}$, Poynting vector $\mathbf{P}$, and momentum density $\mathbf{p}$ of scattered field are expressed as below,

$$\mathbf{E} = \mathbf{E}_{a_2} + \mathbf{E}_{b_2}$$

$$\mathbf{H} = \mathbf{H}_{a_2} + \mathbf{H}_{b_2}$$

$$\mathbf{s} = \frac{\varepsilon}{2k\omega} E_0^2 |a_2|^2 \left[ \frac{75}{4} \left( \frac{\cos\theta + \cos 2\theta}{k^2 r^3} + \frac{3\cos\theta}{k^4 r^5} + \frac{9\cos\theta}{k^6 r^7} \right) \hat{\mathbf{e}}_\phi \right],$$

$$\mathbf{P} = \frac{k}{2\omega\mu} E_0^2 |a_2|^2 \frac{25}{4} \left[ \left( \frac{1 + 2\cos\theta \cos 2\theta + \cos\theta \cos 3\theta}{k^2 r^2} + \frac{9\cos\theta \cos 2\theta}{k^6 r^6} + \frac{18\cos 3\theta + 18\cos\theta \cos 2\theta}{k^8 r^8} \right) \hat{\mathbf{e}}_r + \left( \frac{9\sin\theta \sin 2\theta}{k^6 r^6} + \frac{54\sin\theta \sin 2\theta}{k^8 r^8} \right) \hat{\mathbf{e}}_\theta \right],$$

$$\mathbf{p} = \frac{\varepsilon k}{2\omega} E_0^2 |a_2|^2 \frac{25}{4} \left[ \left( \frac{1 + 2\cos\theta \cos 2\theta + \cos\theta \cos 3\theta}{k^2 r^2} + \frac{9\cos\theta \cos 2\theta}{k^6 r^6} + \frac{18\cos 3\theta + 18\cos\theta \cos 2\theta}{k^8 r^8} \right) \hat{\mathbf{e}}_r + \left( \frac{9\sin\theta \sin 2\theta}{k^6 r^6} + \frac{54\sin\theta \sin 2\theta}{k^8 r^8} \right) \hat{\mathbf{e}}_\theta \right].$$

### A8. $\pi/2$-phase combination of electric and magnetic quadrupole

For Mie coefficient $b_2 = \mathrm{i}\, a_2$, the electric field $\mathbf{E}$, magnetic field $\mathbf{H}$, SAM density $\mathbf{s}$, Poynting vector $\mathbf{P}$, and momentum density $\mathbf{p}$ of scattered field are expressed as below,

$$\mathbf{E} = \mathbf{E}_{a_2} + \mathbf{E}_{b_2}$$

$$\mathbf{H} = \mathbf{H}_{a_2} + \mathbf{H}_{b_2}$$

$$\mathbf{s} = \frac{\varepsilon}{2k\omega} E_0^2 |a_2|^2 \left( \frac{25}{4} \frac{(1 + 2\cos 2\theta)\sin^2\theta \sin 2\phi}{kr^2} \hat{\mathbf{e}}_r + \frac{75}{8} \frac{\sin 4\theta}{k^2 r^3} \hat{\mathbf{e}}_\phi \right),$$

$$\mathbf{P} = \frac{k}{2\omega\mu} E_0^2 |a_2|^2 \left( \frac{25}{8} \frac{2 + \cos 2\theta + \cos 4\theta}{k^2 r^2} \hat{\mathbf{e}}_r + \frac{75}{4} \frac{\cos\theta \sin 2\theta \cos 2\phi}{k^3 r^3} \hat{\mathbf{e}}_\theta - \frac{75}{8} \frac{\sin 4\theta \sin 2\phi}{k^3 r^3} \hat{\mathbf{e}}_\phi \right),$$

$$\mathbf{p} = \frac{\varepsilon k}{2\omega} E_0^2 |a_2|^2 \left( \frac{25}{8} \frac{2 + \cos 2\theta + \cos 4\theta}{k^2 r^2} \hat{\mathbf{e}}_r + \frac{75}{4} \frac{\cos\theta \sin 2\theta \cos 2\phi}{k^3 r^3} \hat{\mathbf{e}}_\theta - \frac{75}{8} \frac{\sin 4\theta \sin 2\phi}{k^3 r^3} \hat{\mathbf{e}}_\phi \right).$$

## B. Spin distribution of scattering field of simple multipoles

We particularly studied the spin contributions from combined $a_1, b_1$, and combined $a_1, a_2$, as they are significantly larger than the other Mie coefficients for the nanoparticles used in our experiment. The *x*-polarized incident light propagates along the *z*-direction and is scattered by a spherical particle, as shown in Fig. S8A. The SAM density obtained at the location (marked as a black dot in Fig. S8A, 10$\lambda$ away from the scatter) changes with the phase and amplitude difference of $a_1, b_1$ and $a_1, a_2$, resulting in a spin locking effect phase diagram shown in Fig. S8B. Notably, the SAM is strong if there is a $\pm\pi/2$ phase difference between the two coupled modes, and disappears if the phase difference approach 0 or π. Therefore, the spin locking effect is generalized from Janus dipole (Eq. S3) to other combinations of two modes with a phase difference. This will be derived in the next section. The highlighted points in Fig. S8B are typical combinations of two given modes with different phases and amplitudes. Their spin textures around the scatter ($r = 10\lambda$) are represented in Fig. S8C, where we sketched a light radiation cone for every case. These scatterings are divided into three different types. One, for instance, the electric dipole, represents topological-insulator-like spin textures with two orbital spin distributions perpendicular to the radiation cones ($a_1, b_1, a_2 = 1, 0, 0$). This is a typical transverse spin. For the Janus dipole, the spins are in parallel to the radiation cones ($a_1, b_1, a_2 = 1, i, 0$), that is, longitudinal spin. In general, the spin from scattering is neither perpendicular nor parallel to the momentum. This property can be studied by calculating the spin as a function of propagating distance away from the particle. As shown in Fig. S9A, for electric dipole, electric quadrupole and Huygens dipole, the SAM of scattered field is always perpendicular to $\hat{e}_r$. For Janus dipole, the SAM is approximately perpendicular to $\hat{e}_r$ in the near field, and as the radial distance increases, the SAM tends to be gradually aligned with $\hat{e}_r$. This is because for the Janus dipole, $\mathbf{s} \propto \frac{\sin^2\theta \sin 2\phi}{kr^2}\hat{e}_r + \frac{\sin 2\theta}{k^2 r^3}\hat{e}_\phi$. Additionally, we investigate the variation of radial component of SAM with respect to the radial distance, as shown in Fig. S9B. When the coupling phase between different multipoles is 0, the radial component of the SAM rapidly decays and vanishing in the far field. However, when the coupling phase is $\pi/2$, it can propagate to the far field with a constant spin.

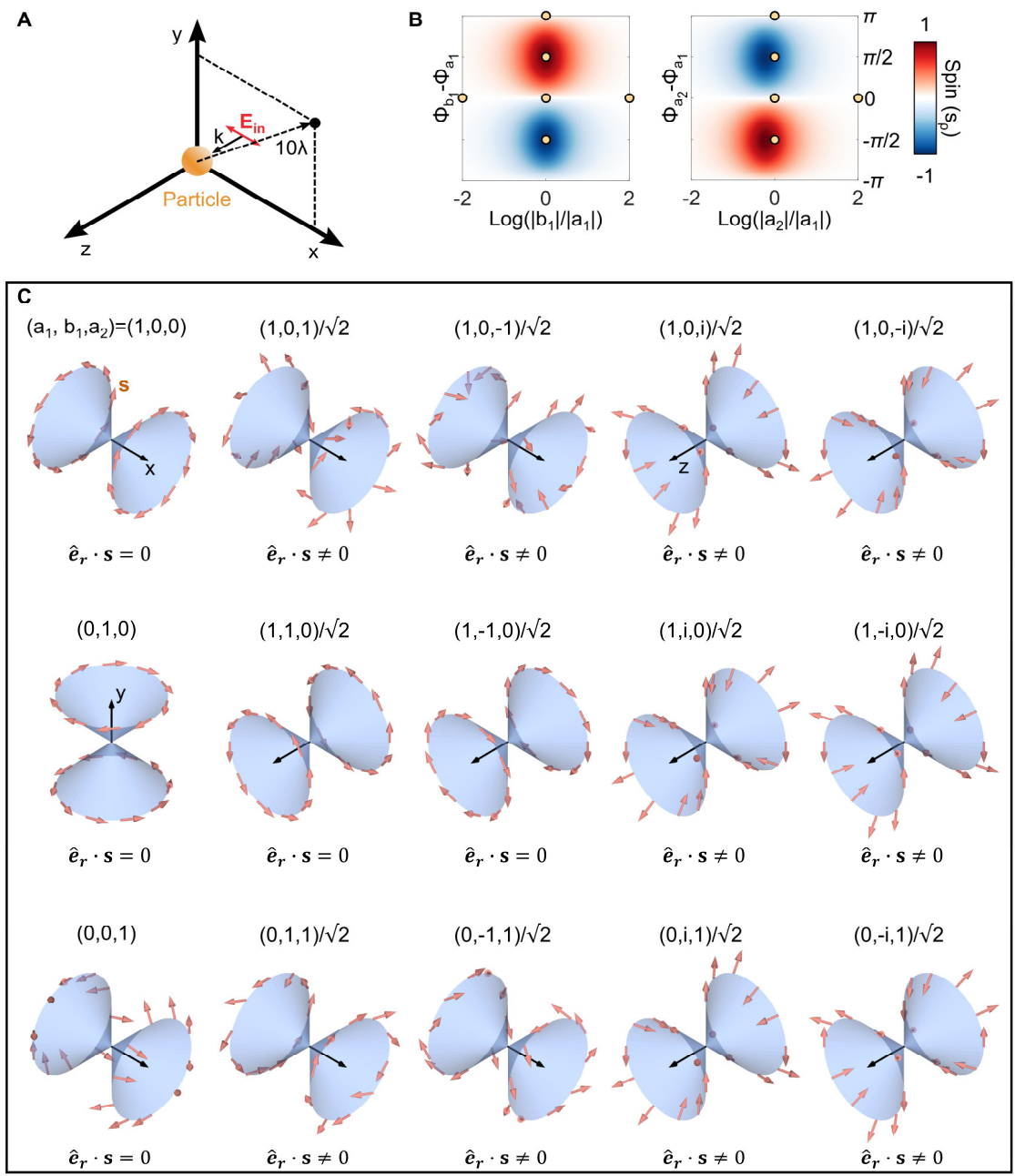

**Fig. S8 SAM properties of scattering field of different combinations of multipoles.** **(A)** Schematic of Mie scattering. The location of the black dot in the figure is $(5\sqrt{2}\lambda, 5\sqrt{2}\lambda, 0)$ **(B)** Phase diagram of $s_\rho$ at the black dot in (A) with respect to Mie scattering coefficients. $s_\rho$ represents radial component of SAM density in $xy$ plane. $\Phi_{a_1}, \Phi_{b_1}, \Phi_{a_2}$ represent the phases of $a_1, b_1, a_2$, respectively. **(C)** SAM distribution of the Mie scattering field under different combinations of Mie coefficients. The red arrows represent the SAM.

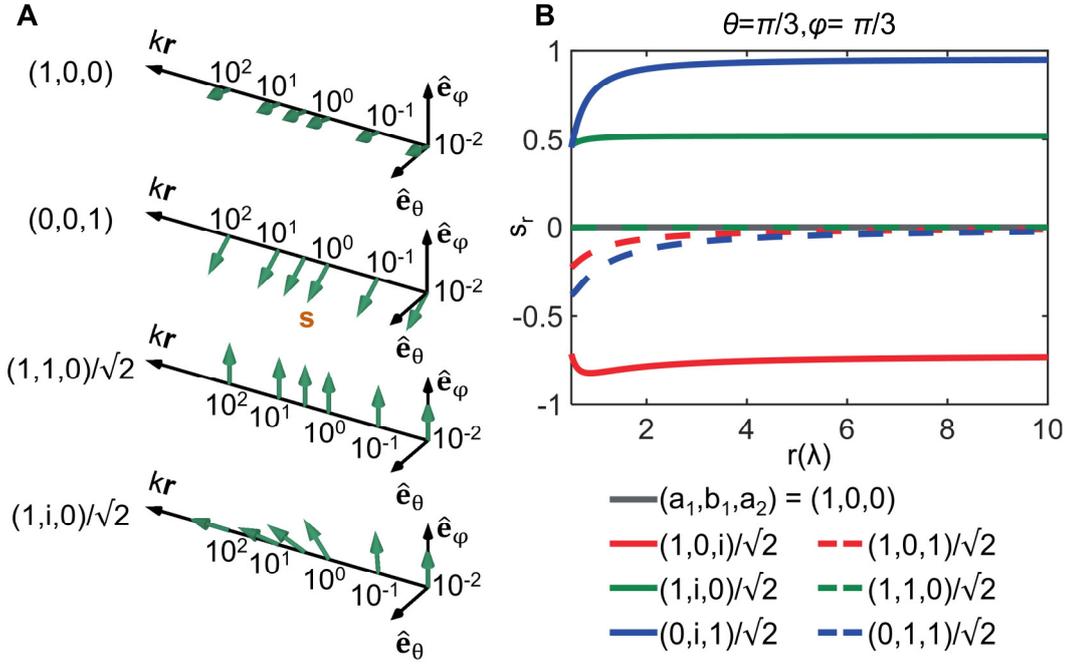

**Fig.S9 The SAM density of scattering field as a function of propagating distance.** **(A)** The change of SAM's direction with respect to $kr$ along the direction of $\theta = \pi/3$, $\varphi = \pi/3$. The green arrows represent the direction of the SAM. **(B)** The normalized radial component of SAM as a function of $r$, which is calculated along the direction of $\theta = \pi/3, \varphi = \pi/3$. $\mathbf{s}_{nor} = \text{Im}[\varepsilon \mathbf{E}^* \times \mathbf{E} + \mu \mathbf{H}^* \times \mathbf{H}]/(\varepsilon|\mathbf{E}|^2 + \mu|\mathbf{H}|^2)$, $s_r = \mathbf{s}_{nor} \cdot \hat{\mathbf{e}}_r$.

## C. SAM density in the far field – the universality of spin effects

From previous theory, we have understood that the Janus dipole, and the electric dipole and quadrupolar combinations, can give rise to far field spin effect akin to a spin Hall effect. Here, we give a theoretical derivation to show the general conditions one requires to obtain far field spin effect from the Mie scattering. Considering $kr \gg 1$, the scattering electric fields are given as (*3*)

$$E_\theta \sim E_0 \frac{e^{i\rho}}{-i\rho} \cos\phi \, S_2(\cos\theta), \text{ and}$$

$$E_\phi \sim -E_0 \frac{e^{i\rho}}{-i\rho} \sin\phi \, S_1(\cos\theta),$$

where $S_1 = \sum_n \frac{(2n+1)}{n(n+1)}(a_n \pi_n + b_n \tau_n)$, $S_2 = \sum_n \frac{(2n+1)}{n(n+1)}(a_n \tau_n + b_n \pi_n)$ and $\pi_n = \frac{P_n^1(\cos\theta)}{\sin\theta}$, $\tau_n = \frac{dP_n^1(\cos\theta)}{d\theta}$, $\rho = kr$. ($\rho$ is different from $\hat{\rho}$ in maintext)

The electric component of SAM density is given by

$$\mathbf{s}_e \propto \text{Im}[\mathbf{E}^* \times \mathbf{E}] = \text{Im}[S_1^* S_2 - S_1 S_2^*] E_0^2 \sin\phi \cos\phi \frac{1}{\rho^2} \hat{\mathbf{e}}_r$$

$$= \sum_{m,n} \frac{(2m+1)}{m(m+1)} \cdot \frac{(2n+1)}{n(n+1)}$$

$$\cdot \begin{bmatrix} -2\pi_m \tau_n |a_m||a_n| \sin(\varphi_{a_m} - \varphi_{a_n}) - 2\pi_n \tau_m |b_m||b_n| \sin(\varphi_{b_m} - \varphi_{b_n}) \\ -2\pi_m \pi_n |a_m||b_n| \sin(\varphi_{a_m} - \varphi_{b_n}) - 2\tau_m \tau_n |a_n||b_m| \sin(\varphi_{a_n} - \varphi_{b_m}) \end{bmatrix} E_0^2 \sin\phi \cos\phi \frac{1}{\rho^2} \hat{\mathbf{e}}_r$$

(S4)

From Eq. (S4), we will see that if there is a non-zero or none-$\pi$ phase difference between any two different multipole components, the SAM will propagate to the far field as a radial component.

## D. The geometric phase from scattering matrix

To theoretically demonstrate that the spin-locking phenomena can be detected from *x*-axis direction in our experiment, we used the far-field scattering matrix to analyze the relationship between the incident electric field and the electric field in the perpendicular direction in the Mie scattering of a single particle. It is well-understood that the spin split phenomena of light usually are interpreted by a geometric phase effect. Mostly, this geometric phase effect is achieve by using rotated anisotropic structures (*13*), or curved optical paths (*14*). We will show that in our experiments, the geometric phase naturally arises from the scattered 3D electric field by the nanoparticles. As mentioned before, the scattered far field is

$$\mathbf{E}_s = E_0 \frac{e^{i\rho}}{-i\rho} \cos\phi\, S_2 \hat{\mathbf{e}}_\theta - E_0 \frac{e^{i\rho}}{-i\rho} \sin\phi\, S_1 \hat{\mathbf{e}}_\phi = E_0 \frac{e^{i\rho}}{-i\rho} [(S_2 \cos\theta \cos^2\phi + S_1 \sin^2\phi)\, \hat{\mathbf{e}}_x + (S_2 \cos\theta \cos\phi \sin\phi - S_1 \cos\phi \sin\phi)\, \hat{\mathbf{e}}_y - S_2 \sin\theta \cos\phi\, \hat{\mathbf{e}}_z].$$

The scattering field is related to the incident field through a scattering matrix **M**, with

$$\mathbf{E}_s = \mathbf{M} \cdot \mathbf{E}_0, \text{ and}$$

$$\mathbf{E}_0 = \begin{bmatrix} E_x \\ E_y \\ 0 \end{bmatrix}.$$

Based on the relation $\mathbf{E}_s = E_0 \frac{e^{i\rho}}{-i\rho} \cos\phi \, S_2 \hat{e}_\theta - E_0 \frac{e^{i\rho}}{-i\rho} \sin\phi \, S_1 \hat{e}_\phi$ and using the rotational symmetry condition $\mathbf{E}_s(\phi; x-\text{polarized}) = \mathbf{E}_s\left(\phi + \frac{\pi}{2}; y-\text{polarized}\right)$, where $x$- or $y$-polarized denotes the polarization of the incident light, we can derive a specific form of the scattering matrix:

$$\begin{bmatrix} E_{sy} \\ E_{sz} \end{bmatrix} = \frac{e^{ik(r-z)}}{-ikr} \begin{bmatrix} \frac{1}{2} S_2 \cos\theta \sin 2\phi - \frac{1}{2} S_1 \sin 2\phi & S_2 \cos\theta \sin^2\phi + S_1 \cos^2\phi \\ -S_2 \sin\theta \cos\phi & -S_2 \sin\theta \sin\phi \end{bmatrix} \begin{bmatrix} E_x \\ E_y \end{bmatrix}. \quad (S5)$$

Therefore, considering the $xy$ plane as the incident plane, and the $yz$ plane as the output plane, there are crossed polarizations in the output plane, akin to those happen in anisotropic structures. The off-diagonal elements in the matrix of Eq. (S5) indicate the coupling between $E_x$ and $E_y$, which is the geometric-phase origin of the spin-locking effect. Particularly, the dependence of the scattering matrix on the angular coordinate $\phi$ results in the spin split observed in our experiment. According to our paraxial optical imaging system, we have $\theta \to \pi/2$ and $\phi \to 0$, and Eq. (5) is simplified as

$$\begin{bmatrix} E_{sy} \\ E_{sz} \end{bmatrix} = \frac{e^{ik(r-z)}}{-ikr} \begin{bmatrix} -S_1\phi & S_1 \\ -S_2 & -S_2\phi \end{bmatrix} \begin{bmatrix} E_x \\ E_y \end{bmatrix}.$$

Let us define $\alpha_1 = \frac{S_1 + S_2}{2}, \alpha_2 = \frac{S_1 - S_2}{2}$, then

$$\begin{bmatrix} -S_1\phi & S_1 \\ -S_2 & -S_2\phi \end{bmatrix} = \alpha_1 \begin{bmatrix} -\phi & 1 \\ -1 & -\phi \end{bmatrix} + \alpha_2 \begin{bmatrix} -\phi & 1 \\ 1 & \phi \end{bmatrix} \quad (S6)$$

The Eq. (S6) describes a universal geometric phase effect in scattering from $\alpha_2$. While $\alpha_1$ and $\alpha_2$ are determined by the size and refractive index of the particle, the geometric phase spatial profile is determined only by the incident wave and observation direction, regardless of the properties of particle. This can be also seen from Fig. S11C, the four-lobe $s_\rho$ distribution from different sizes.

## Section S9. SAM density distribution of a focused beam

We give a comparison between our BSLE and the transverse spin from focused beams by introducing the theory of the latter (*15*). The electric and magnetic field of a focused are

$$\mathbf{E} = \frac{\hat{e}_x + m\hat{e}_y - \frac{x+my}{q(z)}\hat{e}_z}{\sqrt{1+|m|^2}} A(\rho, z)\exp(ikz),$$

$$\mathbf{H} = \frac{k}{\omega\mu} \frac{\hat{e}_y - m\hat{e}_x - \frac{y-mx}{q(z)}\hat{e}_z}{\sqrt{1+|m|^2}} A(\rho, z)\exp(ikz).$$

Here, $A(\rho, z) = A_0 \frac{z_R}{q(z)} \exp\left[ik \frac{\rho^2}{2q(z)}\right]$, $q(z) = z - iz_R$, $z_R = k\omega_0^2/2$, $\rho = \sqrt{x^2 + y^2}$.

The SAM density is given by

$$\mathbf{s} = \mathrm{Im}\left[\frac{\varepsilon}{4\omega}\mathbf{E}^* \times \mathbf{E} + \frac{\mu}{4\omega}\mathbf{H}^* \times \mathbf{H}\right] = \frac{\varepsilon}{2\omega}|A(\rho,z)|^2 \left[\sigma\hat{e}_z + \frac{1}{|q(z)|^2}(\sigma xz - yz_R)\hat{e}_x + \frac{1}{|q(z)|^2}(\sigma yz + xz_R)\hat{e}_y\right].$$

Here, $\tau = \frac{1-|m|^2}{1+|m|^2}$, $\chi = \frac{2\mathrm{Re}m}{1+|m|^2}$, $\sigma = \frac{2\mathrm{Im}m}{1+|m|^2}$ are Stokes parameters.

Considering that the incident light is linearly polarized, namely, $\sigma = 0$, the SAM density is

$$\mathbf{s} = \frac{\varepsilon}{2\omega} \frac{|A(\rho,z)|^2 z_R}{|q(z)|^2}\left[-y\hat{e}_x + x\hat{e}_y\right].$$

The SAM density is a chiral texture in the $xy$ plane, locked with the incident $\mathbf{k}$ following a right-handed rule. Therefore, the spin distribution is independent to the polarization direction of the focused beam (Fig. S10), which is different from our experimental results shown in main text Fig. 4 and Fig. S4.

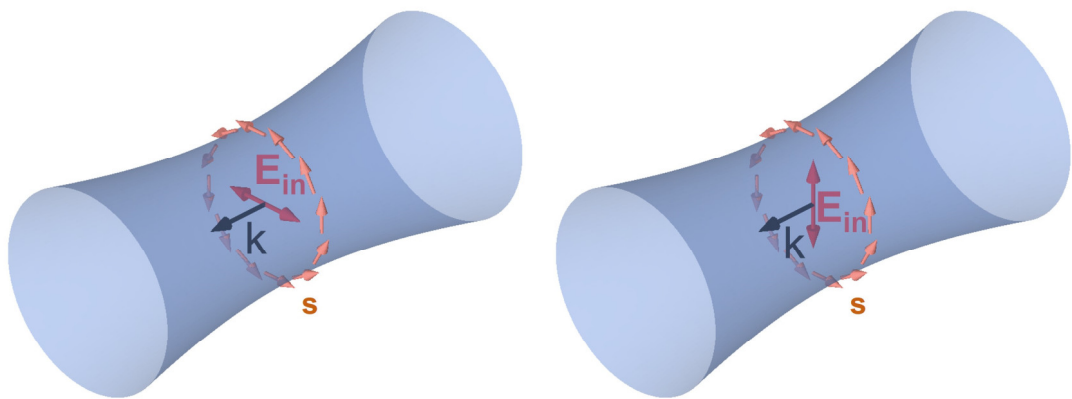

**Fig. S10 SAM density distribution of a focused beam.** The SAM (orange) is produced by the linearly polarized light along the *x* direction **(A)** and *y* direction **(B)**, respectively. Here the black arrow indicates the direction of the wave vector **k**.

## Section S10. The spin from different particle diameter

Figure S11 shows the calculated Mie scattering coefficients $|a_n|$, $|b_n|$ and their phase with the diameter of an AuNP sphere from 20-600 nm (incident wavelength 639 nm). When the diameter is smaller than 50 nm, $|a_2|, |a_3|, |a_4|, |b_1|, |b_2|, |b_3|, |b_4| \ll |a_1|$ and the spin density is perpendicular to the $\hat{e}_r$, $s \propto 1/r^3$. In this region, the scattering is dominated by electric dipole. When the diameter is between 50 and 250 nm, both electric quadrupole and magnetic dipole arise (Fig. S11A), with none-0 and none-π phases between each other (Fig. S11B). According to Eq. (S5), we can see the spin-locking effect in agreement with the experimental observations. The increasement of the size of particle can make the spin pattern more complicated in the *yz* plane (Fig. S11C) but the *xy* plane always remain a four-lobe spin distribution, as predicted by Eq. (S6).

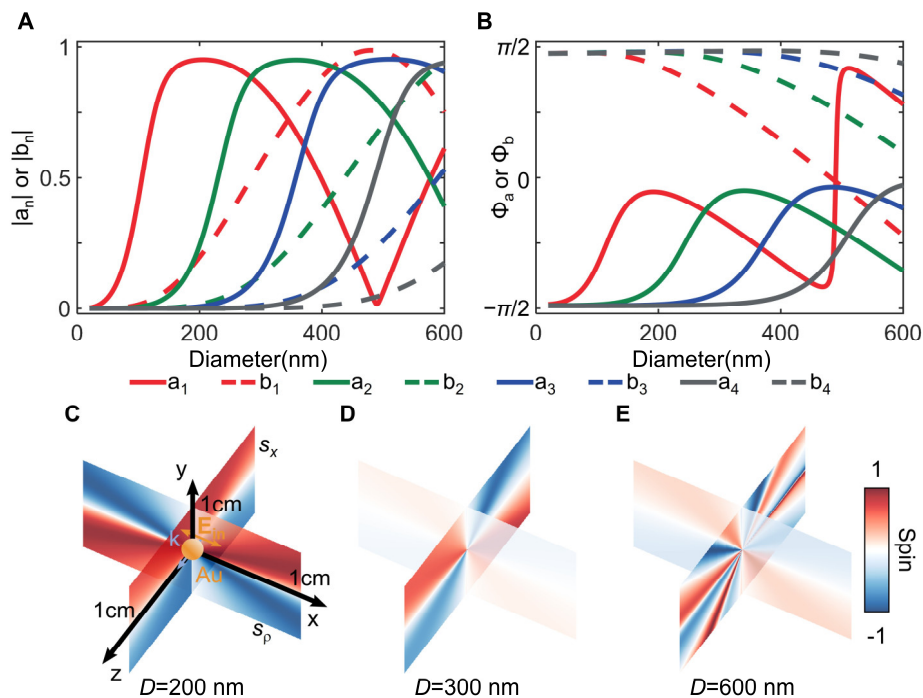

**Fig. S11 Mie coefficients, their phases, and SAM density distribution for AuNP of different sizes. (A)** Calculated Mie coefficients, **(B)** their phases for a single AuNP as a function of diameter. Solid and dotted lines indicate electric and magnetic components respectively. SAM density distribution in the *xy* plane and *yz* plane of a 200 nm-sized AuNP **(C)**, 300 nm-sized AuNP **(D)** and 600 nm-sized AuNP **(E)**.

## Section S11. Dynamic Light Scattering (DLS) theory

In parallel with our BSLE, we used a dynamic light scattering device (Litesizer 500 from Anton Paar) and scanning electron microscope (SEM, Zeiss Ultra Plus Field Emission Scanning Electron Microscope) to characterize the particle size of our samples and compared these results with our optical spin-resolved method.

DLS is based on the Brownian motion of scattering particles in a suspension or a liquid solution. When a single wavelength laser illuminates the particles, the laser is scattered to all directions. The scattering intensity changes with time due to the random movement of the particles. Particularly, smaller particles diffuse faster, causing more rapid fluctuations in the intensity. This random fluctuation of light intensity can be used to extract a diffusion coefficient and calculate the particle size.

The intensity temporal autocorrelation function $g^2(\tau) = \langle I(t) \times I(t+\tau)\rangle / \langle I^2(t)\rangle$ is related to the electric field correlation function $g^1(\tau)$ by the Siegert relation $g^2(\tau) = 1 + \beta|g^1(\tau)|^2$ (*16*), where $\beta$ is a factor related to the geometry and collimation of the laser, $I(t)$ is the intensity of light, and $\tau$ is the time lag value. The coherence time $\tau_c$ of the scattered light can be derived from $g^1(\tau) = \exp(-\Gamma\tau) = \exp(-\tau/\tau_c)$, with $\Gamma = D_t q^2$ and $q = (4\pi n/\lambda)\sin(\theta/2)$ is the wave vector. $\lambda = 658$ nm is the wavelength of the incident laser; $n$ is the refractive index of the sample; and $\theta$ is the angle between the detector and the sample. Hydrodynamic radius was calculated according to the autocorrelation analysis of scattered light intensity data based on translation diffusion coefficient by the Stokes-Einstein relation (*17*):

$$D_h = k_B T / 3\pi\eta D_t,$$

where $D_h$, $D_t$, $k_B$, $T = 298\ K$ and $\eta = 0.89\ mP_a$ are the hydrodynamic diameter, translational diffusion coefficient, Boltzmann's constant, thermodynamic temperature, and viscosity of the water, respectively.

The AuNPs and Fe$_3$O$_4$ particles were hosted in 1-cm-optical-path quartz cuvettes after diluting. The particle size distribution and mean diameter were measured by a DLS device with three fixed scattering angles of 15°, 90° and 175° respectively, at the temperature of 25°C. The results for AuNPs and Fe$_3$O$_4$ are shown in Fig. S12 and Fig. S13, respectively. Meanwhile, AuNPs and Fe$_3$O$_4$ particles were also characterized with SEM (Fig. S12 and Fig. S13), as a good comparison with the results from DLS measurements and our optical spin-resolved spectroscopy.

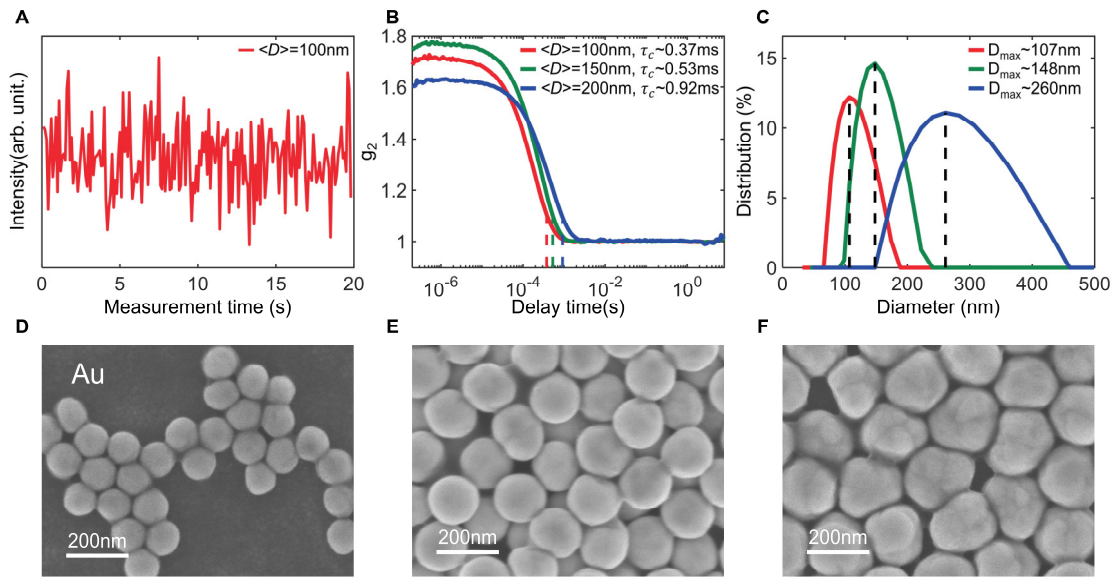

**Fig. S12** (**A**) Measured $I(t)$ by a DLS device, (**B**) autocorrelation function $g^2(\tau)$ and (**C**) size distributions of AuNPs. (**D-F**) SEM images of AuNPs corresponding to mean diameters of 100, 150 nm and 200 nm, from left to right, respectively.

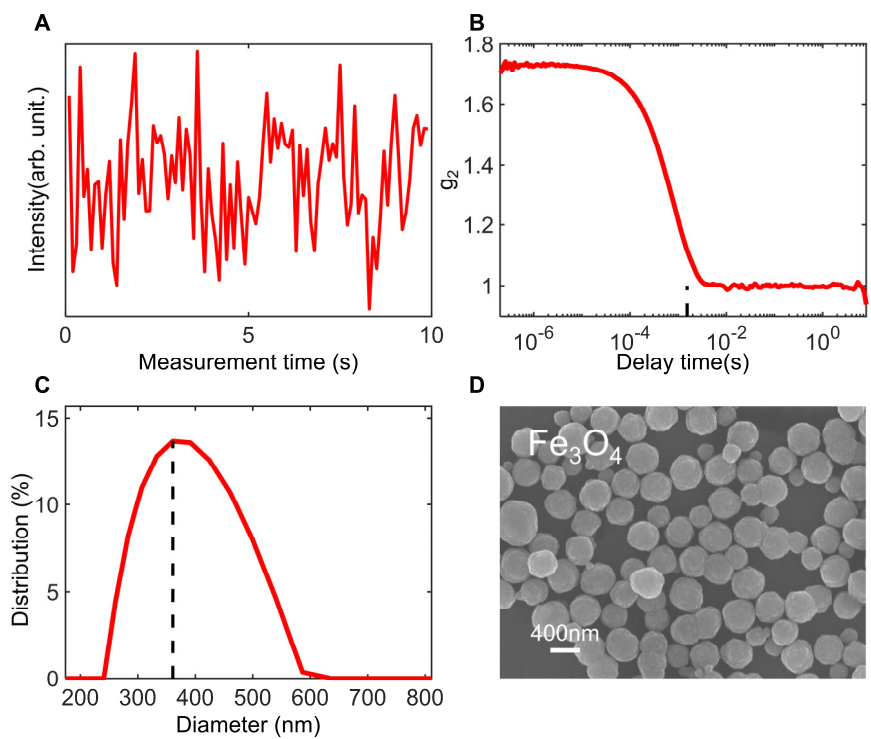

**Fig. S13** (**A**) Measured $I(t)$ by a DLS device, (**B**) autocorrelation function $g^2(\tau)$ and (**C**) size distribution of three monodisperse $Fe_3O_4$ particles. (**D**) SEM images of $Fe_3O_4$ particles with maximum diameter of 360.8 nm.

# Section S12. Evaluate the experimentally obtained spin-resolved spectra

We give detailed calculation method for obtaining the results in main text Figure 5. We randomly placed $N_p = 10^4$ particles in a region of 2 mm × 2 mm × 2 cm, with all the particles the same size, $D$. We defined an observation region centered at $x = 0$ mm, $y = 2.7$ mm and $z = 5$ mm with a perturbation region of 1.2 mm × 1.2 mm in the $yz$ plane. This region is set to simulate the size of a pinhole. Then, we randomly select $N_m = 300$ observation points within this region and calculate the left-handed circularly polarized light intensity with $\bar{I}_L(\lambda) = 0.5 \, \text{abs}(E_y + iE_z)^2$ and the right-handed circularly polarized light intensity with $\bar{I}_R(\lambda) = 0.5 \, \text{abs}(E_y - iE_z)^2$. After averaging the intensities over the 300 observation points to obtain the average intensity distributions $\bar{I}_L(\lambda)$ and $\bar{I}_R(\lambda)$, we calculated the spin distribution $s_x(\lambda) = (\bar{I}_R(\lambda) - \bar{I}_L(\lambda))/(\bar{I}_R(\lambda) + \bar{I}_L(\lambda))$. We set the particle diameters' range from 50 nm to 450 nm, with a 5 nm interval, to obtain a series of spin density spectral curves, as shown in Fig. 5D. Note that the number of $N_p$ and $N_m$ are chosen in a balance of calculation time and stability. Similar results can be observed with different numbers.

To validate the effectiveness of our numerical simulation, we calculate the root mean square error (RMSE) between the experimental spin spectrum and a series of theoretical curves. The difference between the theoretical and experimental results arises from the varying concentrations of nanoparticles – which is a depolarization process making the value of the theory spin different from our observed spin. To take this process into consideration, we introduce a depolarization parameter $dpol$ (from 0 to 1) to calibrate the spin between the experiment and the simulation. Therefore, once we have obtained an experimental spin spectrum, we can use a particle size $D$ and a depolarization parameter $dpol$ to compare the theory and experiment. The agreement between the theoretical and experimental spin spectra is characterized by RMSE =

$\sqrt{\frac{1}{n}\sum_{i=1}^{n}(x_i^{exp}-x_i^{theo})^2}$, where $x_i^{exp}=s_x^{exp}(\lambda_i)$, $x_i^{theo}=dpol \times s_x^{theo}(\lambda_i)$, $dpol$ is related to the concentration and $n$ represent the number of wavelength points in experimental spin spectrum. After substituting these parameters, we have

$$\text{RMSE}=\sqrt{\frac{1}{n}\sum_{i=1}^{n}\left(s_x^{exp}(\lambda_i)-dpol \times s_x^{theo}(\lambda_i)\right)^2}.$$

A smaller RMSE indicates a better agreement between the theoretical and experimental result. The change of RMSE with respect to $dpol$ and theoretical diameter of particle is shown in Fig. S14. The Correlation is equal to 1 − RMSE as shown in Fig. 5B. For AuNPs with a particle diameter of <D> = 100 nm used in the experiment, the best theoretical corresponding diameter of particle is 100 nm. At this point, $dpol=1$, RMSE ≈ 0.022. For AuNPs with a particle diameter of <D> = 150 nm used in the experiment, the best theoretical corresponding diameter of particle is 135 nm. At this point, $dpol=0.79$, RMSE ≈ 0.032. For AuNPs with a particle diameter of <D> = 200 nm used in the experiment, the best theoretical corresponding diameter of particle is 195 nm. At this point, $dpol=0.23$, RMSE ≈ 0.032.

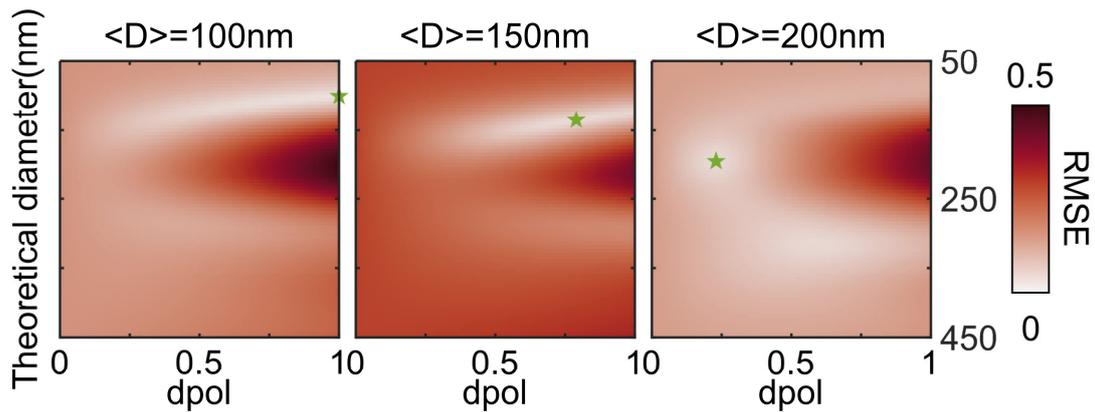

**Fig. S14 RMSE distribution with respect to $dpol$ and theoretical diameter of particle.** The value of RMSE is minimal at the location marked by the green star. The corresponding $dpol$ and theoretical diameter are used to calculate Fig. 5D.

# Section S13. Statistics of the scattering from incoherent to coherent region

As mentioned in the main text, we establish a two-part scattering theory to evaluate the complex scattering of light in this Brownian system. This process is sketched in Fig. S15. (i) A two-step scattering process for each combination of two particles: one in the ballistic region and another in the diffusion region. The light field of the incident laser, $E_0$, excites one of the nanoparticles in the ballistic region $r_b$, resulting in a radiation source, $E_b = M \cdot E_0$. Here, $M$ is Mie scattering operator. In addition, another particle in the diffusion region $r_d$ is similarly excited by $E_b$, resulting in $E_d = M \cdot E_b$, and approximately $E_d \propto E_b(r')$, with $r' = r_d - r_b$. (ii) The field superposition of $E_d$ from many particles with different $r_b$.

For incoherent scattering, we first project the field $E_b(r')$ onto the spin-up and spin-down states, obtaining the opposite spin intensities $I_R(r')$ and $I_L(r')$, respectively. The intensities are summed over all particles, and $I_{R,L}(r_d) = \sum_{r_b} I_{R,L}(r_d - r_b)$. For coherent case, we directly sum all field $E_b(r')$ from ballistic particles and the electric field $E(r_d) = \sum_{r_b} E_b(r_d - r_b)$, and then projecting this field into spin-up and spin-down states to have the opposite spin intensities $I_R(r_d)$ and $I_L(r_d)$. In general, the scattering theory can be developed to partial coherence (*18*). Here, we statistically divide the system composed of a number of $N$ particles into $m$ equal parts. Within each part, the scattering from these particles are coherent, but between different parts, the fields from each other are completely incoherent. This simple approach provides us an artificial parameter characterized by $m/N \in [0,1]$, through which we can calculate a clear evolution of the scattering property from incoherent to coherent. The scattering within the same group is coherent, and the electric field is superposed as: $E_m = \sum_n E_{mn}$. Scattering between different groups is incoherent, and the light intensity is superposed as: $I_R = 0.5 \sum_m \text{abs}(E_{my} - iE_{mz})^2$, $I_L = 0.5 \sum_m \text{abs}(E_{my} + iE_{mz})^2$. The spin is then calculated as $s_x = (I_R - I_L)/(I_R + I_L)$. This process is repeated $10^4$ times to avoid statistical error. The solid curves shown in Fig. 3D are fitted by a Burr distribution for the intensity and Beta distributions for the spin. By changing $m/N$ from 100% to 0.01%, the corresponding curves are shown in Fig. 3D from left to right, which indicates that the most significant spin split phenomenon occurs in the incoherent scenario. The reduce of $m/N$ will reduce this effect until the phenomenon disappears in a coherent disordered system. With respect to reduction of $m/N$, the absolute value of the skewness in main text Fig. 3D (spin panel) is about 0.056 for *m/N* = 100%, 0.433 for *m/N* = 0.08%, 0.639 for *m/N* = 0.02%, 0.610 for *m/N* = 0.01%. The skewness of reddish curve is positive, and of blueish curve is negative.

Appendix: Beta Distribution (*19, 20*)

Probability density function (PDF):

$$f(x) = \frac{x^{\alpha-1}[1-x]^{\beta-1}}{B(\alpha,\beta)}, 0 \leq x \leq 1.$$

Here, $\alpha$, $\beta$ are shape parameters and $B(\alpha,\beta)$ is the normalization constant. Skewness is a measurement of the asymmetry of the probability distribution of a real-valued random variable relative to its mean. The skewness of a Beta distribution is given by

$$\gamma_1 = \frac{2(\beta-\alpha)\sqrt{\alpha+\beta+1}}{(\alpha+\beta+2)\sqrt{\alpha\beta}}.$$

When $\alpha < \beta$ and $\gamma_1 > 0$, the distribution is concentrated on the left side of the graph with a longer right tail. When $\alpha > \beta$ and $\gamma_1 < 0$, the distribution is concentrated on the right side of the graph with a longer left tail. Since the spin is in a region of $[-1, 1]$, we use $(s_x + 1)/2$ to map the spin value from $[-1, 1]$ to $[0, 1]$. After fitting with Beta distribution, in the upper diffusion region the skewness is $\gamma_1 < 0$, with the tail on the left side of the distribution (Fig. 3D, spin statistics). For the lower diffusion region, the skewness is $\gamma_1 > 0$, with the tail on the right side of the distribution.

Burr Type XII distribution (*19, 20*)

Probability density function (PDF):

$$f(x) = \frac{ck}{\lambda} \frac{(x/\lambda)^{c-1}}{(1+(x/\lambda)^c)^{k+1}}, x > 0$$

Here, $c$, $k$, $\lambda$ are all positive numbers. $\lambda$ is a scale parameter, $c$, $k$ are shape parameters, where $k$ only affects the characteristics of the right tail, while $c$ influences the characteristics of both tails. The normalized light intensity values lie within $[0, 1]$, fitted by Burr Type XII distribution (main text Fig. 3D).

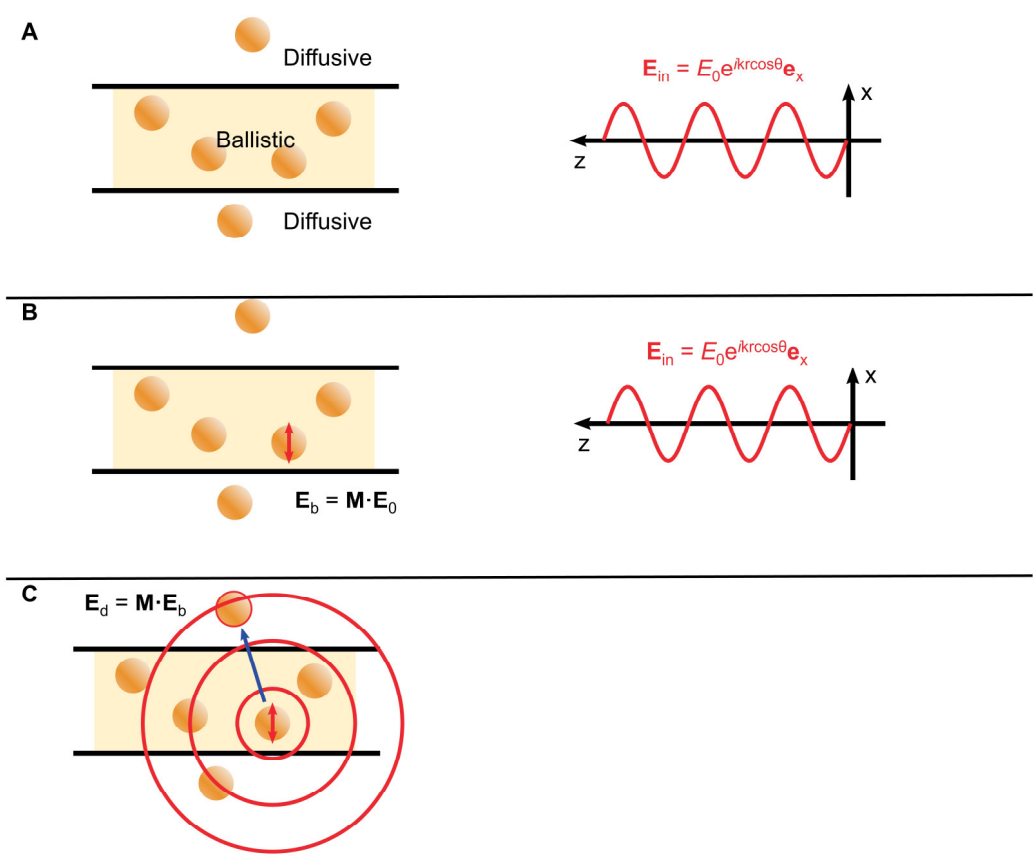

**Fig. S15 Sketched essential scattering processes. (A)** Schematic of scattering system. **(B)** First step: The light field of the incident laser, $\mathbf{E}_0$, excites one of the nanoparticles in the ballistic region, resulting in a radiation source $\mathbf{E}_b = \mathbf{M} \cdot \mathbf{E}_0$. Here, $\mathbf{M}$ represents the Mie scattering matrix. **(C)** Second step: Another nanoparticle in the diffusion region is similarly excited by $\mathbf{E}_b$, resulting in $\mathbf{E}_d = \mathbf{M} \cdot \mathbf{E}_b$.

# Supplementary References